\documentclass[12pt]{JHEP}
\usepackage{amsmath,amssymb}
\usepackage{ae,aecompl,cancel,comment}

\usepackage{subfigure}
\usepackage{yfonts}
\usepackage{pdfsync}
\usepackage{graphicx}

\newcommand{\be}{\begin{equation}}
\newcommand{\ee}{\end{equation}}
\newcommand{\beqa}{\begin{eqnarray}}
\newcommand{\eeqa}{\end{eqnarray}}

\def\bemat{\left( \begin{array}}
\def\enmat{\end{array} \right)}

\relax

\preprint{MPP-2011-154}

\title{Magnetic catalysis with massive dynamical flavours}

\author{Johanna Erdmenger \footnotemark[1]
\\Max-Planck-Institut f\"{u}r Physik (Werner-Heisenberg-Institut)
\\ F\"{o}hringer Ring 6, 80805 M\"{u}nchen, Germany}

\author{Veselin Filev \footnotemark[2]
\\Max-Planck-Institut f\"{u}r Physik (Werner-Heisenberg-Institut)
\\ F\"{o}hringer Ring 6, 80805 M\"{u}nchen, Germany
\vspace{.1cm}
\\\hspace{3.1cm}\&
\\School of Theoretical Physics
\\Dublin Institute for Advanced Studies
\\10 Burlington Road, Dublin 4, Ireland}

\author{Dimitrios Zoakos \footnotemark[3]
\\Centro de F\'\i sica do Porto \& Departamento de F\'\i sica e Astronomia
\\Faculdade de Ci\^encias da Universidade do Porto
\\Rua do Campo Alegre 687, 4169--007 Porto, Portugal}

\footnotetext[1]{E-mail address: \email{jke@mppmu.mpg.de}}
\footnotetext[2]{E-mail address: \email{filev@mppmu.mpg.de}}
\footnotetext[3]{E-mail address: \email{dimitrios.zoakos@fc.up.pt}}

\abstract{Within gauge/gravity duality, we construct a backreacted supergravity background dual to SU($N_c$) ${\cal N} = 4$ SYM coupled to $N_f$ massive fundamental flavours 
in the presence of an external magnetic field. Our solution is perturbative in a parameter that counts the number of the internal flavour loops. The background has a hollow cavity in the bulk of the geometry, where it is similar to the supergravity dual of a 
non-commutative SYM. The radius of this cavity is related to the dynamically generated mass of the fundamental fields.
We apply our construction to study the effect of magnetic catalysis and develop an appropriate renormalization scheme to compute the free energy 
and the fundamental condensate of the  dual gauge theory as a function
of the bare mass. While at leading order in the expansion of the 
perturbative parameter,  
the free energy and the fundamental condensate agree with the results obtained in the quenched approximation, at next order we show that the effect of magnetic 
catalysis is enhanced and the contribution to the condensate of the theory from internal fundamental loops runs logarithmically with the finite cutoff~$\Lambda_{UV}$.}
 
\begin{document}

%%%%%%%%%%%%%%%%%%%%%%%%%%%%%%%%%%%%%%%%%%%%%%%%%%%%%%%%%%%%%%%%%%%
\newpage
\section{Introduction}

In recent years, research on 
the AdS/CFT correspondence \cite{Maldacena:1997re} and its
applications has expanded significantly, leading to substantial new
results. At present, holographic descriptions of non-perturbative phenomena range 
from applications to condensed matter systems (e.g superconductivity,
superfluidity, quantum Hall effect) to applications in high energy
physics and to the quark-gluon plasma (e.g.~confinement/deconfinement
phase transitions, chiral symmetry breaking, elliptic flow,
hadronization). Despite the remarkable insights gained by studying
holographic gauge theories, the application of the correspondence to
phenomenologically relevant gauge theories continues to remain
a challenge. A way
to further improve the phenomenological relevance of the AdS/CFT
correspondence is to explore further phenomena of universal
nature. The results of such studies are expected to capture
aspects of the qualitative behaviour of realistic gauge theories (such as QCD), 
which do not possess holographically dual supergravity backgrounds.

\paragraph{}

An important example in this class of phenomena is the effect of mass generation and spontaneous chiral symmetry breaking in the presence of  an external magnetic field. 
This effect has been shown to be model-independent and therefore insensitive to the microscopic physics underlying the low energy effective theory. Its essence is the dimensional 
reduction $D \to D-2$,  (3+1 $\to$ 1+1) in the dynamics of fermion
pairing in a magnetic field. The enhanced infrared divergences in
lower dimensions suggests that the dynamics of fermion pairing is
governed by the lowest Landau level, which hints at the universal nature of the phenomenon. Furthermore, one can show that the $(D-2)$-dimensional dynamics of the Landau level favours the condensation of the fermion pairs. This effect is known as Magnetic Catalysis. Magnetic Catalysis has been demonstrated in various (1+2)- and (1+3)-dimensional field theories \cite{Gusynin:1994re}-\cite{Klimenko:1992ch} using conventional field theoretical methods. The holographic approach to this effect has been initiated in \cite{Filev:2007gb}, where the (1+3)-dimensional holographic gauge theory dual to the D3/D7--brane intersection has been considered\footnote{For a comprehensive review we refer the reader to \cite{Filev:2010pm}.}. Additional holographic studies of magnetic catalysis at finite temperature or chemical potential for both (1+3)- and (1+2)-dimensional systems have been performed in \cite{Filev:2007qu}-\cite{Bolognesi:2011un}. 

\paragraph{}

In the holographic description of magnetic catalysis, the flavour
degrees of freedom are introduced by an additional stack of flavour
D--branes. The most understood and widely applied regime of this
approach is when the flavour branes are in the probe limit and their
backreaction to the ambient supersymmetric background is
neglected~\cite{Karch:2002sh}. On the field theory side, this
corresponds to the ``quenched" approximation in which the number of
flavour degrees of freedom, $N_f$, is much smaller than the number of
colour degrees of freedom, $N_c$. In terms of Feynman diagrams this
implies ignoring the contribution from internal quark loops (windows)
in the planar diagrams of the corresponding large $N_c$ expansion. At
present all holographic studies of magnetic catalysis are in the
quenched approximation. An obvious question is to ask how corrections
due to internal quark loops would influence the effect of magnetic catalysis. Our goal is to provide such an estimate in a perturbative expansion in the ratio between the number of flavour and colour degrees of freedom, $N_f/N_c$.

\paragraph{}

For obtaining an unquenched holographic description of magnetic
catalysis, we have to take into account the backreaction of the
flavour branes to the supergravity background sourced by the colour
branes. Ideally such a background would describe localized branes,
however for technical reasons  this is a difficult task  even in the
supersymmetric case. To circumvent these difficulties, the flavour
D--branes may be distributed along the compact directions of the
supergravity background. This procedure is called
smearing\footnote{For a detailed review on the smearing see
  \cite{arXiv:1002.1088} while for other solutions employing this
  technique see \cite{allsusyunquenched}.}. The smearing restores a significant part of the global symmetry of the geometry and hence simplifies the corresponding Einstein equations. 

\paragraph{}

Supersymmetric backgrounds of smeared massless flavour D--branes have been constructed in \cite{hep-th/0612118},
and for flavours with finite bare mass in \cite{arXiv:0807.0298}.  In the case of massive flavours, these backgrounds display a hollow cavity in the bulk of the geometry, where the supergravity solution is sourced solely by the colour branes. The radius of this cavity is related to the bare mass of the fundamental flavors. In the limit of vanishing bare mass the cavity shrinks to the radius of the compact part of the geometry and the supergravity background has an essential singularity at the origin of the non-compact part of the geometry. In both cases the dilaton field diverges at large radial distances. This corresponds to the Landau pole that the dual field theory develops in the UV, due to its positive beta function $\beta\propto N_f/N_c$.

\paragraph{} 

We can imagine that a non-supersymmetric background interpolating
between two supersymmetric backgrounds, corresponding to massless
flavors in the UV and massive flavors in the IR, would describe dynamical mass generation. The radius of the hollow cavity then corresponds to the dynamically generated constituent mass of the fundamental flavors. 
A promising framework for the construction of such a geometry was developed in \cite{arXiv:0909.2865}, where the ten-dimensional black-hole solution dual
to the non-conformal plasma of flavoured ${\cal N} = 4$ supersymmetric Yang-Mills theory is presented\footnote{All the hydrodynamic transport coefficients of the model were analyzed in \cite{hydro}, while the addition of a finite baryon density was presented in \cite{Bigazzi:2011it}. 
For a review on unquenching the Quark Gluon Plasma see \cite{Bigazzi:2011db}. }. 
The authors outline the smearing procedure, 
derive the corresponding equations of motion and present a perturbative solution for general massless non-supersymmetric flavour D7--brane embeddings.

\paragraph{}

The first steps towards unquenching the holographic description of magnetic catalysis have been undertaken in \cite{arXiv:1106.1330}, where the authors, following the approach of  \cite{arXiv:0909.2865}, construct a perturbative non-supersymmetric background with a non-vanishing
$B$-field, which corresponds to an external magnetic field coupled to the fundamental degrees of freedom of the dual gauge theory. 
In the case of massless fundamental fields and sufficiently strong magnetic field, the supergravity background is unstable, suggesting that the theory undergoes a phase transition to a stable phase with dynamically generated mass for the matter fields.

%%%%%%%%%Outline of the paper%%%%%%%%%%%%%

\paragraph{}

In section 2 of this paper we complete the studies initiated in \cite{arXiv:1106.1330} constructing a perturbative non-supersymmetric background with a non-trivial $B$-field, 
for massive flavour fields. Our solution has a hollow cavity in the bulk of the geometry where it is very similar to the supergravity dual of a non-commutative supersymmetric Yang-Mills theory, \cite{Maldacena:1999mh} \& \cite{Hashimoto:1999ut}. The difference is in the presence of a squashed $S^5$, instead of a non-squashed one, 
which breaks the supersymmetry. As suggested above, the radius of this cavity, $r_q$, is related to the dynamically generated mass of the fundamental fields.

\paragraph{}

For radial distances greater than $r_q$ the solution is characterized by a non-vanishing density for the smeared D7--brane charge. At sufficiently large radial distance our solution approaches the supesymmetric one, constructed in \cite{arXiv:0807.0298}. Following the prescription of \cite{arXiv:0909.2865}, we introduce an additional large radial parameter $r_*\gg r_q$ (corresponding to a finite UV cutoff ), at which we match our solution to the  supersymmetric one. Furthermore, we identify the value of the $B$-field at $r_*$ as the magnetic field of the dual gauge theory, $H_*\equiv B(r_*)$. 

\paragraph{}

For radial distances greater than $r_*$ the supergravity background is well approximated by the non-perturbative supersymmetric background \cite{arXiv:0807.0298}. This enables us to relate non-perturbatively the UV parameters of the theory, namely the finite cutoff $\Lambda_{UV}\propto r_*$ and the energy scale corresponding to the landau pole of the theory $\Lambda_{\rm LP}\propto r_{\rm LP}$, where $r_{\rm LP}$ is the radial distance at which the dilaton field diverges. 

\paragraph{}

Our supergravity construction has the following renormalization group flow interpretation: 

\paragraph{}

At the energy scale set by the finite cutoff ($\Lambda_{UV}\propto r_*$) the dual gauge theory is a commutative ${\cal N}=4$ Supersymmetric Yang-Mills theory coupled to $N_f$ flavours of ${\cal N}=2$ hypermultiplet fundamental fields. The fundamental hypermultiplets are coupled to a constant external magnetic field $H_*$, which breaks the supersymmetry.
Decreasing the energy scale the Yang-Mills theory becomes non-commutative and the parameter of non-commutativity (roughly the non-trivial part of the $B$-fleld) is proportional to the ratio $N_f/N_c$. 

\paragraph{}

At energy scales of the order of the physical mass of the fundamental fields, $M_q$ (roughly $M_q\sim r_q$) the flavour fields decouple (the D7--brane charge density vanishes). At lower energy scales (inside the cavity, $r<r_q$) the dual gauge theory is a pure (only adjoint degrees of freedom) non-commutative Yang-Mills theory. To leading order the parameter of non-commutativity in the plane perpendicular to the magnetic field scales as $\Theta^{23}\propto \frac{N_f}{N_c}\frac{1}{H_*}$. 

\paragraph{}

Note that in the bare Lagrangian of the dual gauge theory the external magnetic field $H_*$ couples explicitly only to the fundamental degrees of freedom. Therefore the non-commutativity of the adjoint degrees of freedom cannot be captured by the quenched approximation and is one of the novel results of our analysis.

\paragraph{}

Finally, in section 3 of this work we apply our construction to study the effect of magnetic catalysis. We develop an appropriate renormalization scheme and compute the free energy and the fundamental condensate of the holographically dual gauge theory as a function of the bare mass of the fundamental degrees of freedom. Our studies show that to leading order in a perturbative expansion in the ratio $N_f/N_c$, the free energy and fundamental condensate of the theory agree with the results obtained in the quenched approximation. Furthermore, at next order in  $N_f/N_c$ we show that the effect of magnetic catalysis is enhanced and the contribution to the condensate of the theory from internal fundamental loops runs logarithmically with the finite cutoff $\Lambda_{UV}$.

%%%%%%%%%%%%%%%%%%%%%%%%%%%%%%%%%%%%%%%%%%%%%%%%%%%%%%%%%%%%%%%%%%%%
\section{Constructing the Background}

In the present section we will construct the supergravity background necessary for the holographic study of the phenomenon of magnetic catalysis.
The field theory duals are realized on the intersection between a set of $N_c$ {\it colour} D3-branes and a 
set of $N_f$, homogeneously smeared, {\it flavour} D7--branes, with an additional coupling between the fundamental fields and an external magnetic field.
The {\it colour} D3-branes are placed at the tip of a Calabi-Yau (CY) cone over a Sasaki-Einstein manifold $X_5$, 
where the latter can be expressed as a $U(1)$ fiber bundle over a four-dimensional K\"ahler-Einstein base (KE).
The {\it flavour} D7--branes extend along the radial direction, wrap a submanifold $X_3$ of $X_5$ and smear  
homogeneously over the transverse space \cite{hep-th/0602027,hep-th/0505140}.

%%%%%%%%%%%%%%%%%%%%%%%%%%%%%%%%%%%%%%%%%%%%%%%%%%%%%%%%%%%%%%%%

\subsection{Ansatz \& smearing of the flavours}

To take into account the contribution from internal fundamental loops in the Veneziano limit of the dual gauge theory, we need to consider the backreaction of the flavour branes. Ideally the corresponding supergravity solution would describe localized D7--branes which break the global symmetry of the internal subspace of the supergravity background from $SO(6)$ down to $SO(4)\times SO(2)$ (when $X_5$ is $S^5$). However, even in the supersymmetric case (no external magnetic field), this is a very difficult task. One way to circumnavigate the technical difficulties is to construct a solution with smeared D7--branes.

In general the smearing procedure involves distributing the branes at different locations in the transverse space subspace and consider a course grained approximation in which the sum over all individual embeddings becomes an integral over a distribution of branes. It is somewhat analogous to the smearing of point-like charges in electrostatics in 1+3 dimensions to obtain: one, two or three dimensional charge densities.

In our case the smearing is performed in such a way that the isometries of the fibered K\"ahler-Einstein space are kept unbroken,  
allowing for an ansatz where all the unknown functions just depend on a  single radial coordinate. 
Based on this assumption we adopt the following ansatz for the metric
\begin{equation}
ds_{10}^2 = h^{-\frac{1}{2}}\left[-dt^2 + dx_1^2+b(dx_2^2+dx_3^2)\right] + h^\frac{1}{2}
\left[ b^2 S^8F^2 d\sigma^2 + S^2 ds_{CP^2}^2 + F^2 (d\tau + A_{CP^2})^2 \right] \, ,
\label{10dmetric}
\end{equation}
where the $CP^2$ metric is given by 
\begin{eqnarray}
ds_{CP^2}^2&=&\frac{1}{4} d\chi^2+ \frac{1}{4} \cos^2 \frac{\chi}{2} (d\theta^2 +
\sin^2 \theta d\varphi^2) + \frac{1}{4} \cos^2 \frac{\chi}{2} \sin^2 \frac{\chi}{2}(d\psi + \cos \theta d\varphi)^2
\quad \& \nonumber \\
A_{CP^2}&=& \frac12\cos^2 \frac{\chi}{2}(d\psi + \cos \theta d\varphi)\,\,.
\label{cp2metric}
\end{eqnarray}
The range of the angles is $0\leq (\chi, \theta) \leq \pi$,  $0\leq \varphi, \tau < 2\pi$, $0\leq \psi< 4 \pi$.
The ansatz for the NSNS and the RR field strengths is given by
\begin{eqnarray} \label{NS+RR}
& B_{2}  =  H dx^2\wedge dx^3 \, , \quad
C_{2} = J \,dt \wedge dx^1 \,,  & 
\nonumber \\ \nonumber \\
& F_{5}  =  Q_c\,(1\,+\,*)\varepsilon(S^5)\, , \quad
F_{1} = Q_f \, p(\sigma) \,(d\tau + A_{CP^2})\, , \quad
F_{3} = d C_2\,+\, B_{2} \wedge F_1 \, , &
\end{eqnarray}
where $\varepsilon(S_5)$ is the volume element of the internal space\footnote{with $\int \, \varepsilon(S_5) \, = \, \text{Vol}(S^5) \, = \, \pi^3$ } 
and $Q_{c}, Q_{f}$ are related to
the number of different colours and flavours in the following way
\begin{equation}
N_c = \frac{Q_c\, Vol(X_5)}{(2\pi)^4g_s \,\alpha'^2} \quad \& \quad
N_f = \frac{4\,Q_f\,Vol(X_5)}{Vol(X_3) g_s} \, .\label{QcQf}
\end{equation}
In our case $X_5=S^5$ and the volume of the three sphere is $2\pi^2$. All the functions that appear in the ansatz, $h,b,S,F, \Phi, J \, \& \, H$,
depend on the radial variable $\sigma$ only. In the convention we follow, $S \, \& \, F$ have dimensions of length, 
$p,b,h,J \, \& \, H$ are dimensionless and $\sigma$ has a dimension of length${}^{-4}$. Furthermore, $\sigma=\infty$ at the origin and decreases to $\sigma_*$ at the boundary.

The function $b$ in the ansatz for the metric reflects the breaking of the $SO(1,3)$ Lorentz symmetry down to $SO(1,1)\times SO(2)$. The function $p(\sigma)$ in $F_{(1)}$ and $F_{(3)}$, determines the distribution of the brane embeddings and has a characteristic asymptotic behavior. In fact  $p(\sigma)$ encodes the bare mass, $m_q$ and the fundamental condensate of the dual gauge theory. It vanishes at energy scales smaller than the quarks' mass while it asymptotes to $1$ in the UV. This leads to the formation of a spherical cavity inside the bulk of geometry. The radius of this cavity sets the energy scale related to the physical mass of the quark, $M_q$\footnote{We remind the reader that beyond the quenched approximation the dual field theory has a positive $\beta$-function. As a result the mass of the fundamental fields runs with the energy scale and even in the supersymmetric case the physical mass $M_q$ differs from the bare mass $m_q$. } . 

We can understand better the structure of the distribution function $p(\sigma)$ if we consider a representative D7--brane embedding, the so called ``fiducial embedding". The fiducial embedding is an auxiliary  D7--brane embedding, which probes the backreacted geometry, its shape determines the distribution function $p(\sigma)$.

\begin{figure}[h] %  figure placement: here, top, bottom, or page
   \centering
   \subfigure[A couple of supersymmetric flat embeddings of mass $m_q=M_q\sim r_q$. Note that in the quenched approximation there is no mass generation.]{ \includegraphics[width=3in]{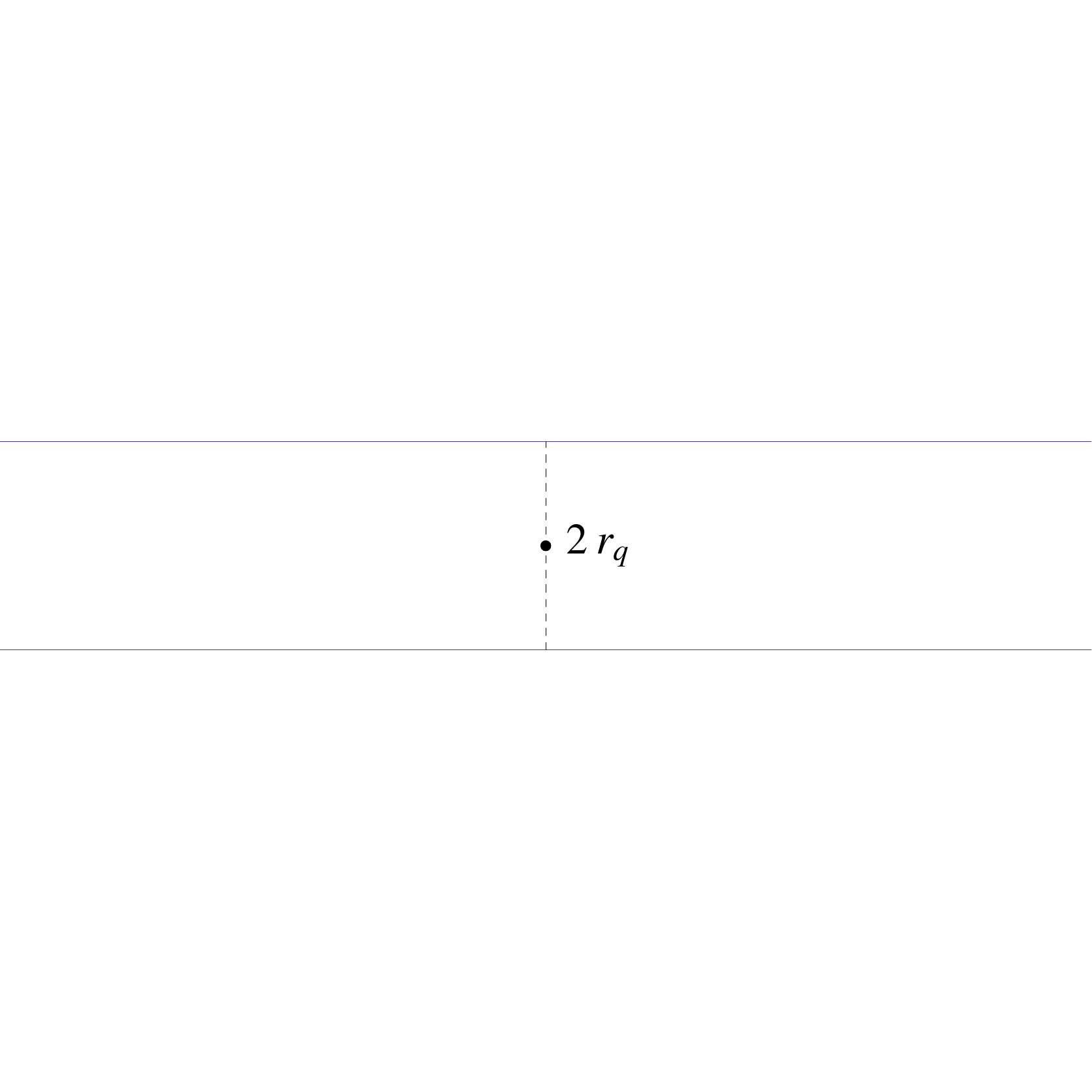} \label{fig:fig(-1)a}} 
  \quad\quad\quad   
    \subfigure[A couple of non-supersymmetric mass generating embeddings of vanishing bare mass $m_q=0$ and physical mass $M_q\sim r_q.$]{ \includegraphics[width=3in]{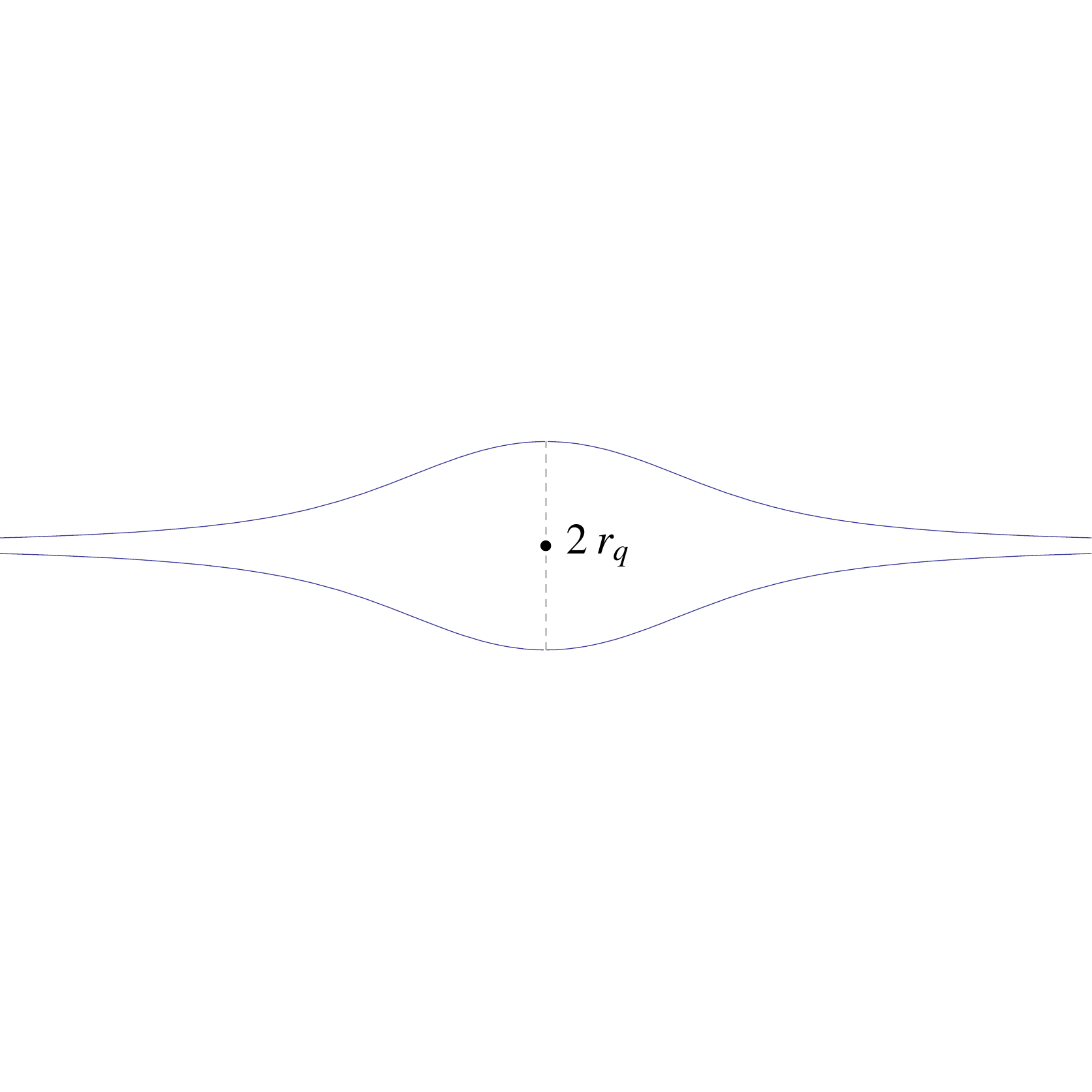} \label{fig:fig(-1)b}} \\
    \subfigure[Smeared supersymmetric embeddings.]{ \includegraphics[width=3in]{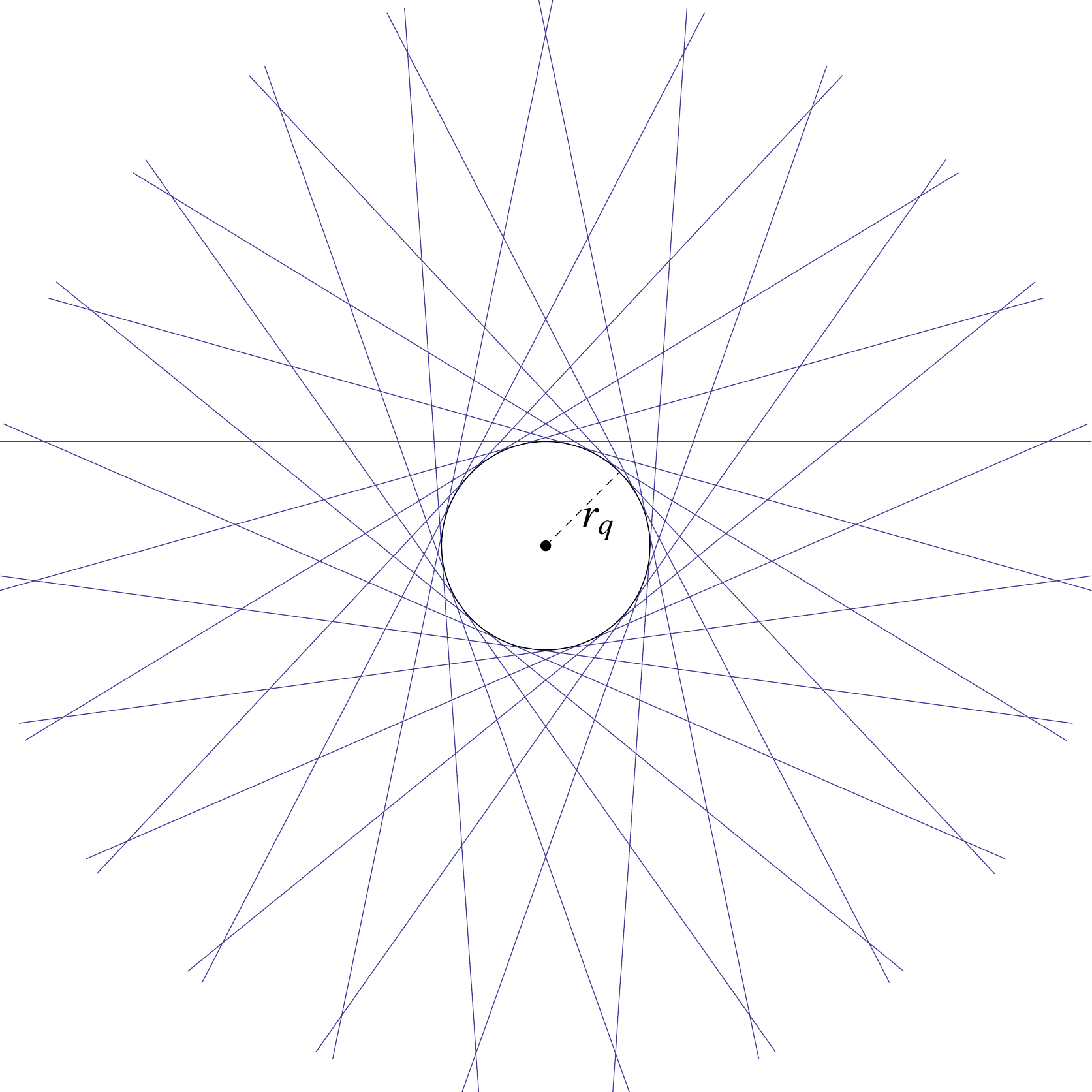} \label{fig:fig(-1)c}}
   \quad\quad\quad   
   \subfigure[Smeared mass-generating embeddings.]{ \includegraphics[width=3in]{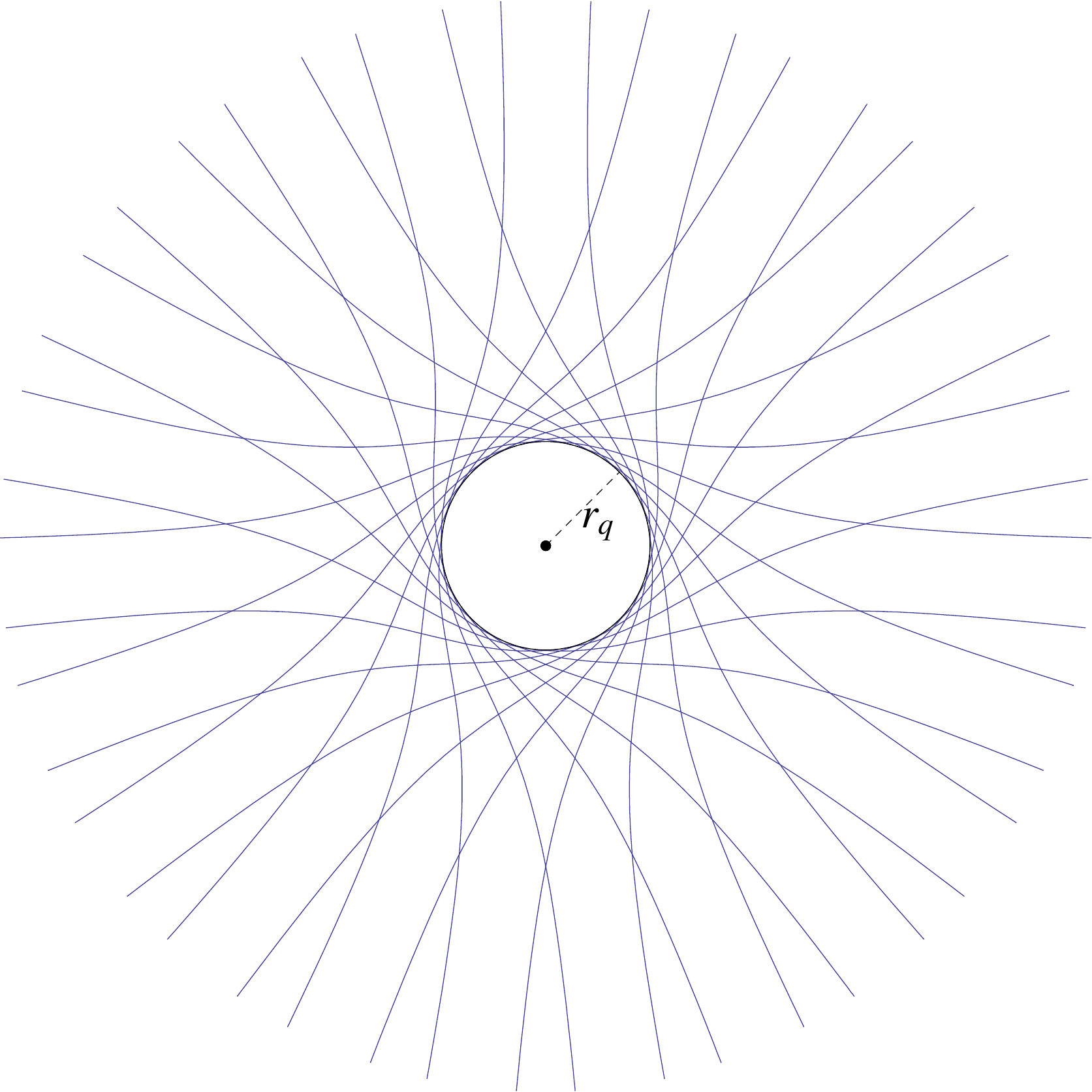} \label{fig:fig(-1)d}}
   \caption{A visualization of the smearing procedure.}
   \label{fig:fig(-1)}
\end{figure}

 To leading order in a multipole expansion we can describe the profile of the fiducial embedding by functions, which depend only on the radial (holographic) coordinate $\sigma$. The fiducial embedding relevant for our study wraps an internal three-cycle parameterized by  $\theta,\varphi,\psi$, extends along $\sigma$ and sits at a fixed value of  $\tau$. To obtain the distribution function $p(\sigma)$ we smear the fiducial embedding by acting with the symmetries of the internal space (see appendix of \cite{arXiv:0909.2865}). In this way we obtain 
\begin{equation}
p(\sigma)=\cos^4\frac{\chi_{q}}{2}\, .
\label{p-sigma}
\end{equation}
To clarify the relationship between the distribution function and the profile of the fiducial embedding (\ref{p-sigma}), in figure \ref{fig:fig(-1)} we have presented plots of a family of D7--brane embeddings  obtained by acting on the fiducial one with a discrete subgroup of the symmetry group of the internal space. Figures \ref{fig:fig(-1)a} and \ref{fig:fig(-1)b} represent probe D7--brane embeddings terminating at the same minimal radial distance $r_q$. The first one is a supersymmetric embedding (vanishing magnetic field), while the second one is a non-supersymmetric one (finite magnetic field) and exhibits mass generation. Figures \ref{fig:fig(-1)c} and \ref{fig:fig(-1)d} represent the smearing of the corresponding probe D7--brane embeddings. One can clearly see the formation of a spherical cavity in the bulk of the geometry. On can also compare the radial distributions corresponding to supersymmetric and non-supersymmetric fiducial embeddings\footnote{Note that we have taken advantage of the fact that to leading order in the ratio $N_f/N_c$ the profile of the fiducial embeddings is well approximated by the profile of the probe embeddings.}.

\newpage
In the next subsection following the strategy initiated in \cite{arXiv:0909.2865} we will describe the whole system in terms 
of a one-dimensional effective action, from which all the equations of motion can be produced. In principle the same system of equations
can be explicitly derived also from a ten-dimensional point of view after inserting the above ansatz in the ten-dimensional equations of motion plus the 
Bianchi identities. 

The equations of motion  and the corresponding Bianchi identities for the NSNS and RR fields are relatively easy to obtain. However, writing down the ten-dimensional Einstein equations is a difficult task. The crucial step is to obtain an effective ten-dimensional expression for the smeared the DBI action, in the case of massive non--supersymmetric probes. While in the massive supersymmetric case such a construction is possible using calibrated geometry (see e.g. \cite{arXiv:1002.1088,arXiv:0811.3646}) in the non-supersymmetric case this is a non-trivial task. To circumvent this difficulty we derive the equations of motion from a one dimensional effective action. The key point is that even if we were able to obtain a simple ten dimensional term for the smeared DBI action we would still adopt the ansatz (\ref{10dmetric})-(\ref{NS+RR}) for the supergravity fields, which is equivalent to reducing the action to one dimension. Therefore we could first reduce the eight dimensional DBI action of every individual flavour brane to one dimension and then sum them. In this context the statement that the flavour branes are smeared is equivalent to the statement that their reduced one dimensional actions are equivalent. In other words, the smearing of the DBI action is equivalent to reducing the eight dimensional DBI action of a fiducial flavour brane embedding to one dimension and multiplying the result by the number of flavour branes $N_f$.

%%%%%%%%%%%%%%%%%%%%%%%%%%%%%%%%%%%%%%%%%%%%%%%%%%%%%%%%%%%%%%%%

\subsection{Effective action \& the equations of motion}

The action for the Type IIB supergravity plus the contribution from the $N_f$ D7--branes in the Einstein frame is 
\begin{equation}
S=S_{IIB} + S_{fl} \, , \label{genact}
\end{equation}
where the relevant terms of the $S_{IIB}$ action are
\begin{eqnarray} \label{TypeIIB action}
S_{IIB}&=&\frac{1}{2\kappa_{10}^2}\int d^{10} x \sqrt{-g} \Bigg[ R
- {1 \over 2} \partial_M\Phi \partial^M\Phi
- {1 \over 2} e^{2\Phi}F_{(1)}^2 
- {1 \over 2} \frac{1}{3!} e^{\Phi}  F_{(3)}^2
- {1 \over 2} \frac{1}{5!} F_{(5)}^2  \nonumber\\ \nonumber\\
&&\qquad \qquad \qquad \qquad \qquad  \quad \qquad \qquad  \,
- {1 \over 2} \frac{1}{3!} e^{-\Phi}  H_{(3)}^2\Bigg] 
- \frac{1}{2\kappa_{10}^2}\, \int C_4\wedge H_3\wedge F_{3} \, ,
\end{eqnarray}
and the action for the flavour D7--branes takes the usual DBI+WZ form
\begin{equation}
S_{fl} = -T_7 \sum_{N_f} \Bigg[ \int d^8x\, e^\Phi 
\sqrt{-\det (\hat{G}+e^{-\Phi/2} {\cal F}}) \, 
- \, \int \left(\hat{C}_8 + \hat{C}_6 \wedge B_2 \right) \Bigg]\,,
\label{actionflav}
\end{equation}
with ${\cal F} \equiv  B + 2 \pi \alpha' F$. In those expressions $B$ denotes a non-constant magnetic field,
$F$ the worldvolume gauge field and the hat refers to the pullback of the quantities, 
along the worldvolume directions of the D7--brane.  The gravitational constant and D7--brane tension, 
in terms of string parameters, are
\begin{equation}
\frac{1}{2\kappa_{10}^2} = \frac{T_7}{g_s} = \frac{1}{(2\pi)^7g_s^2 \alpha'^4} \, .
\end{equation}

The first step in our analysis is deriving the equation of motion for the fiducial embedding,
which follows from the action $S_{fl}=S_{DBI}+S_{WZ}$. The DBI action for the D7--brane is given by
\begin{equation}
S_{DBI} = -\frac{T_7}{8} \int d^8x\, e^{\Phi} \,b^2 \, \sin\theta \, S^6 \, F^2 \, 
\cos^3  \frac{\chi_{q}}{2} \, \Xi_1 \, \, \Xi_2 \, \, \Xi_3 \, ,
\label{DBI}
\end{equation}
where we have introduced the following auxiliary dimensionless quantities
\begin{equation}
\Xi_1 \equiv \sqrt{\cos^2 \frac{\chi_{q}}{2}+
\frac{S^2}{F^2}\sin^2 \frac{\chi_{q}}{2}}\,\,,
\quad
\Xi_2\equiv \sqrt{1+\frac{(\partial_\sigma\chi_{q})^2}{4b^2 \,S^6\,F^2 }}
\quad \& \quad 
\Xi_3\equiv \sqrt{1+\frac{e^{-\Phi}\,H^2\,h}{b^2}}\ ,
\end{equation}
with $\chi_{q}$ being a function of $\sigma$ determining the brane embedding. 
The WZ piece of the action is
\begin{equation}
S_{WZ} = - \frac{T_7}{32} \, Q_f \int d^8x \sin\theta \, p(\sigma) \, b^2 \,e^{2\Phi} \, S^8 \, 
\Xi_3^2 \, \cos^4 \frac{\chi_{q}}{2} \,.
\label{WZ}
\end{equation}
The corresponding equation of motion for $\chi_{q}$ is given by 
\begin{eqnarray}
\label{EOM-chi}
0&=&\frac{1}{2} \partial_\sigma\left[e^\Phi \cos^3 \frac{\chi_{q}}{2}\,
\frac{\Xi_1 \Xi_3}{\Xi_2}\,(\partial_\sigma \chi_{q})
\right]+\\
&+&e^\Phi b^2 \, S^6 F^2 \, \Xi_3 \, \cos^2 \frac{\chi_{q}}{2}\sin \frac{\chi_{q}}{2}
\Bigg[3 \, \Xi_1 \, \Xi_2 + \cos^2 \frac{\chi_{q}}{2} \left(1-\frac{S^2}{F^2}\right) \frac{\Xi_2}{\Xi_1}
+Q_f \, e^\Phi \frac{S^2}{F^2}\cos \frac{\chi_{q}}{2} \, p(\sigma) \, \Xi_3 \Bigg] \, .
\nonumber
\end{eqnarray}
As we commented in the previous subsection all functions depend only on $\sigma$, hence it is possible
to describe the system in terms of a one-dimensional effective action. Inserting all the ingredients in 
\eqref{genact} we obtain:
\begin{equation}
S_{eff}=\frac{\pi ^3 V_{1,3}}{2\kappa_{10}^2}\int {\cal L}_{1d} \, d\sigma \label{S-eff-1}\ ,
\end{equation}
where $V_{1,3}$ is the volume of the Minkowski space and ${\cal L}_{1d} $ is given by the 
following expression
\begin{eqnarray} \label{L-effective1}
{\cal L}_{1d} &=&
-\frac{1}{2} \left(\frac{h'}{h}\right)^2 + 12 \left(\frac{S'}{S}\right)^2 + 8 \, \frac{F' S'}{F S}
+ 24\,b^2\, F^2\,S^6 - 4\, b^2\, F^4\,S^4
\nonumber \\ \nonumber \\
&+&\frac{b'}{b}\,\left( \frac{h'}{h}+ 8 \,\frac{S'}{S}+ 2\, \frac{F'}{F} \right)+  
\frac{1}{2}\, \left(\frac{b'}{b}\right)^2 - \frac{b^2 Q_c^2}{2 h^2} 
-\frac{1}{2}\,Q_f^2\, p^2\, b^2e^{2\Phi} S^8 \,\Xi_3^2
\\ \nonumber \\
&-&4\,Q_f \, e^{\Phi}\, b^2\,F^2\,S^6 \, \Xi_1 \, \Xi_2 \, \Xi_3 \, \cos^3 \frac{\chi_{q}}{2}
- \frac{1}{2}\,\Phi'^2
- \frac{1}{2}\,\frac{e^{-\Phi}\,H'^2\,h}{b^2} \left(1 - \frac{e^{2\Phi}\,J'^2\,b^2}{H'^{2}}\right)
-Q_c H J' \, . 
\nonumber 
\end{eqnarray}
Since the potential $J$ enters the effective action only through its derivative, 
it corresponds to a ``constant of motion". This new parameter is
related to the value of the magnetic field close to the boundary through the
equations of motion for $F_3$, coming from the 10d supergravity~\cite{arXiv:1106.1330}. We will fix this
constant of motion in the following way
\begin{equation} \label{defJ}
\frac{\partial {\cal L}_{1d}}{\partial J'} \, \equiv - \, Q_c H_{\star} \quad \Rightarrow \quad 
J' \,=\, \frac{e^{-\Phi}\,Q_c}{h} \left(H-\, H_{\star}\right)\, .
\end{equation}
The next step is to use equation \eqref{defJ} to eliminate
$J'$, in favor of $H_*$, in equation \eqref{L-effective1} after performing the following 
Legendre transformation 
\begin{equation}  \label{L-effective2}
\tilde {\cal L}_{1d} = L_{1d}-\frac{\delta L_{1d}}{\delta J'} \, J'\Bigg|_{J' \equiv J'(H,H_*)} \, ,
\end{equation}
and then calculate the Euler-Lagrange equations from the transformed
action \eqref{L-effective2}. Defining the following auxiliary (dimensionless) expressions 
\begin{eqnarray}
&& \Xi_4 \equiv 1- \cot^2 \frac{\chi_q}{2}\,\, \frac{F^2}{S^2} \, , \quad 
\xi \equiv  \cos^3 \frac{\chi_q}{2}\, \frac{\Xi_1 \,\Xi_2}{\Xi_3}\, , 
\\
&& \beta_2 \equiv 1 + \frac{e^{2\Phi}\,J'^{2}\,b^2}{H'^2} \quad \& \quad
\beta_3 \equiv 1 + \frac{e^{-2\Phi}\,H'^{2}\,\beta_2}{Q_f^2\,p^2\,H^2\, b^2\,S^8} \nonumber\ ,
\end{eqnarray}
we can write the equations of motion in the following compact way
\begin{eqnarray}
\partial_\sigma^2(\log b)&=&-\, 4 Q_f\,H^2\, h \,S^6 \, F^2\, \xi
\,-\,e^{\Phi}\,H^2\,Q_f^2\, p^2\, h \,S^8\,\beta_3 \label{diff-b}\ ,
\\ \nonumber \\
\partial_\sigma^2(\log h)&=&- \, Q_c^2 \frac{b^2}{h^2}\,
-\,\frac{1}{2}\,e^{\Phi}\,H^2\,Q_f^2\,p^2\,h \,S^8\,\beta_3\,
+\,\left(1-\beta_2\right)\,\frac{e^{-\Phi}\,h\,H'^2}{b^2} \label{diff-h}\ ,
\\ \nonumber \\ 
&-& 2 Q_f\,H^2\, h \,S^6 F^2 \xi\,\ ,
\nonumber \\ \nonumber \\ 
\partial_\sigma^2(\log S)&=& -2\, b^2 \, F^4 \, S^4 + 6\, b^2 \, F^2 \, S^6 +
\frac{1}{4} \, Q_f^2 \, p^2 \, e^{\Phi} \, H^2\, h \,S^8\, \beta_3 \label{diff-S}\ ,
\\ \nonumber \\
&-&\frac{1}{2}\,Q_f \, e^\Phi \, b^2 \, F^2 \, S^6\, \cos^3 \frac{\chi_q}{2} \,\, \Xi_3
\Bigg[\cos^2 \frac{\chi_q}{2} \, \frac{\Xi_2}{\Xi_1} \, + \, 
\frac{\Xi_1}{\Xi_2}\left(1\,-\,\Xi^2_2\,+\, 2\frac{\Xi_2^2}{\Xi_3^2} \right)\Bigg]\ ,
\nonumber \\ \nonumber \\
\partial_\sigma^2(\log F)&=& 4\, b^2 \,F^4 \,S^4 - 
\frac{1+\Xi_3^2}{4} \, Q_f^2 \, p^2 \, e^{2\Phi} \, b^2\, S^8\,
+\,\frac{1}{4}\,\frac{e^{-\Phi}\,h\,H'^2\,\beta_2}{b^2} \label{diff-F}\ ,
\\ \nonumber \\ 
&-& 2 \, Q_f \, e^{\Phi} \, b^2\, S^8 \, \cos^3 \frac{\chi_q}{2} \, \sin^2 \frac{\chi_q}{2} \, 
\frac{\Xi_2}{\Xi_1 \Xi_3} \, \Bigg[ 1\, - \, \frac{1}{2}\, \, \Xi_4 \, \left(1 - \Xi_3^2 \right) \Bigg]\ ,
\nonumber \\ \nonumber \\
\partial_\sigma^2\Phi&=& \frac{1+\Xi_3^2}{2}
\Bigg[Q_f^2\,p^2\,e^{2\Phi}\,b^2\,S^8 +  4 Q_f\,b^2\, e^\Phi\,F^2\, S^6
\, \xi \Bigg]
-\frac{1}{2}\,\frac{e^{-\Phi}\,h\,H'^2\,\beta_2}{b^2} \label{diff-Phi}\ ,
\\ \nonumber \\ 
\partial_{\sigma} \left[\frac{e^{-\Phi}\,h\,H'}{b^2}\right]&=& e^{\Phi}\,Q_f^2
\, p^2\,H\,h\,S^8\,+\,Q_c\,J'
+\, 4 Q_f \,H \, h \, S^6 \, F^2 \, \xi \, . \label{diff-H}
\end{eqnarray}
Together with the above system of EOM we get the following ``zero-energy'' constraint
\begin{eqnarray} \label{constraint}
0&=&
-\frac{1}{2} \left(\frac{h'}{h}\right)^2 + 12 \left(\frac{S'}{S}\right)^2 + 8 \, \frac{F' S'}{F S}
- 24\,b^2\, F^2\,S^6 + 4\, b^2\, F^4\,S^4 \,  - \,  \frac{1}{2}\,\Phi'^2
\nonumber \\ \nonumber \\
&+&\frac{b'}{b}\,\left( \frac{h'}{h}+ 8 \,\frac{S'}{S}+ 2\, \frac{F'}{F} \right)+  \frac{1}{2}\, 
\left(\frac{b'}{b}\right)^2 + \frac{b^2 Q_c^2}{2 h^2} \,
- \frac{1}{2}\,\frac{e^{-\Phi}\,H'^2\,h}{b^2} \left(1 - \frac{e^{2\Phi}\,J'^2\,b^2}{ H'^{2}}\right) 
\\ \nonumber \\
&+&\frac{1}{2}\,Q_f^2\, p^2 \, b^2 \, e^{2\Phi} S^8 \, \Xi_3 \, + \, 
4\,Q_f\, b^2\,e^{\Phi}\,F^2\,S^6  \,\frac{ \Xi_1 \, \Xi_3}{\Xi_2 } \, \cos^3 \frac{\chi_{q}}{2} \, .
\nonumber 
\end{eqnarray}
Equation (\ref{constraint}) is obtained by requiring invariance of the one dimensional effective action (\ref{S-eff-1}) under an infinitesimal reparameterization $\sigma\to(1+\delta\lambda)\sigma$. This is equivalent to requiring that the one-dimensional Hamiltonian ${\cal H}_{1d}$ corresponding to ${\cal L}_{1d}$ vanishes. An important observation is that in generating the Hamiltonian ${\cal H}_{1d}$ one should Legendre transform with respect to all fields in ${\cal L}_{1d}$, including the field $\chi_q(\sigma)$ specifying the profile of the fiducial embedding. This suggests that we can treat $\chi_q(\sigma)$ as a dynamical variable, therefore we can also obtain the equation of motion (\ref{EOM-chi}) from varying the effective lagrangian (\ref{L-effective1}). Remarkably if we use equation (\ref{p-sigma}) to substitute for $p(\sigma)$ in the effective Lagrangian ${\cal L}_{1d}$ and derive the corresponding Euler-Lagrange equation for $\chi_q(\sigma)$, we reproduce exactly the equation of motion~(\ref{EOM-chi}). This provides a non-trivial self-consistency check of the construction outlined above. 

The system \eqref{defJ} \& \eqref{diff-b}--\eqref{diff-H} allows for a systematic expansion of all the functions
in power series of $Q_f$, as defined in equation (\ref{QcQf}). In fact physically it is more relevant to expand in the parameter, $\epsilon_*$
\begin{equation}
{\epsilon_*}\equiv Q_f\, e^{\Phi_*}\ ,\label{epsilonstar}
\end{equation}
which takes into account the running of the effective 't Hooft coupling (through the dilaton factor $e^{\Phi_*}$ in \eqref{epsilonstar}). We consider the following first order expansion in $\epsilon_*$:
\begin{eqnarray}
& b \, = \, 1 \, + \, \epsilon_* b_1 \, , \quad 
h \, = \, \frac{R^4}{r^4} \, \left(1+\epsilon_*h_1\right) \, , \quad 
S \, = \, r \, \left(1+\epsilon_*S_1\right)\ , &
\label{expansion1}
\\ \nonumber \\ 
& F\, = \, r \, \left(1+\epsilon_*F_1\right)  \, , \quad
\Phi \, = \, \Phi_*+\epsilon_*\Phi_1 \, , \quad
H \, = \, H_* \left( 1 \, + \, \epsilon_* H_1 \right)  \quad \& \quad 
\chi_q \,  = \,  \chi_0 + \epsilon_* \chi_1 \, ,&
\nonumber
\end{eqnarray}
together with the reparametrization:  
\begin{equation}
r_m^4=e^{-\Phi_*}H_*^2R^4  \quad \& \quad 
\tilde r=\frac{r}{r_m}  \, ,\label{dimless-r}
\end{equation}

where $R^4\equiv Q_c/4$.
The result is a coupled system of second order differential equations which can be decoupled by the transformations: 
\begin{equation}
\Delta_1 \, \equiv \,  F_1 \, - \, S_1 \, ,  \quad \quad
\Lambda_1  \, \equiv \, F_1 \, + \, 4 \, S_1 \, + \, \frac{5}{4} \, \, b_1 \quad \quad \& \quad  \quad
\lambda_1 \, \equiv \, h_1 \, - b_1 \, \ .
\end{equation}
For the decoupled system we obtain:
\begin{eqnarray} \label{EOMs}
&& 
\chi_0'' \, + \, \frac{5}{\tilde{r}} \, \frac{{\tilde r}^4 \,+\,\frac{3}{5}}{{\tilde r}^4\,+\,1} \,\chi_0' 
+\, \frac{{\tilde r} \left({\tilde r}^4 \,+\,\frac{1}{2}\right)}{{\tilde r}^4\,+\,1} \,\chi_0'^{\,3} \, = \,  
\, - \, \frac{6}{{\tilde r}^2} \, 
\tan \frac{\chi_{0}(\tilde{r})}{2} \, \sqrt{1 \, + \, \frac{\tilde{r}^2}{4} \, \chi_{0}'(\tilde{r})^2} \, , 
\nonumber \\ \nonumber \\
&&
\lambda_1'' \, + \, \frac{5}{\tilde{r}} \, \lambda_1' \, -\, \frac{32}{\tilde{r}^2} \, \lambda_1 \, = \,  
\frac{1}{2} \, J_{\lambda_1} \, , 
\qquad \qquad \,\,
H_1'' \, + \, \frac{1}{\tilde{r}} \, H_1' \, -\, \frac{16}{\tilde{r}^2} \, H_1 \, =
\, \frac{1}{2}\, J_{H_1} \, ,
\nonumber \\ \nonumber \\
&& 
\Phi_1'' \, + \, \frac{5}{\tilde{r}} \, \Phi_1' \, = \, \frac{1}{2} \, J_{\Phi_1} \, ,
\qquad \qquad \qquad \qquad \qquad \qquad \quad 
b_1'' \, + \, \frac{5}{\tilde{r}} \, b_1' \, = \, \frac{1}{2} \, J_{b_1} \, , 
\nonumber \\ \nonumber \\
&&
\Delta_1'' \, + \, \frac{5}{\tilde{r}} \, \Delta_1' \, -\, \frac{12}{\tilde{r}^2} \, \Delta_1 \, = 
\, \frac{1}{2}\, J_{\Delta_1} \, ,
\qquad \qquad 
\Lambda_1'' \, + \, \frac{5}{\tilde{r}} \, \Lambda_1' \, -\, \frac{32}{\tilde{r}^2} \, \Lambda_1 \, = 
\, \frac{1}{2}\, J_{\Lambda_1} \, ,
\end{eqnarray}
where the analytic expressions for the sources $J_{\lambda_1}, J_{H_1}, J_{\Phi_1}, J_{b_1}, J_{\Delta_1}, J_{\Lambda_1}$ appear in Appendix \ref{nonsusy-ingredients}. The equation coming from the constraint (\ref{constraint}) is: 
\begin{equation} \label{EOM-eta}
\eta_1' \, - \, \frac{4}{{\tilde r}} \, \eta_1 \, = \, - \, J_{\eta_1} ({\tilde r})
\end{equation}
with $\eta_1 \, \equiv 2 \Lambda_1\, + \, \lambda_1$ and $J_{\eta_1}$ given in Appendix \ref{nonsusy-ingredients}.

As one can see all of the equations of motion, except the one for the fiducial embedding, in \eqref{EOMs} are linear, therefore it is possible to obtain their solution in an integral form in terms of  appropriate Greens functions. 
On the other hand the non-linear equation of motion for $\chi_0$ is the same as the one for a probe D7--brane studied in ref. [ours]. 
In fact it is the only non-linear equation that we have in our construction and we solve it numerically.  
The general solution for the classical D7--brane has the following expansion at large $\tilde r$
\begin{equation}
\sin\frac{\chi_0(\tilde r)}{2}=\frac{\tilde m_0}{\tilde r}+\frac{\tilde c}{\tilde r^3}+O\left(\frac{1}{\tilde r^5}\right)\ ,
\end{equation}
where $\tilde m_0\equiv m_0/r_m$ \&  $\tilde c\equiv c/r_m^3$, while $m_0$ \& $c$ are proportional to the bare mass and fundamental condensate of the dual field theory. Inside the bulk of the geometry $\chi_0$ terminates at a given radial distance $\tilde r_{\rm min}=\tilde r_q$ at which the $S^3$ cycle wrapped by the D7--brane vanishes. 
As described at the beginning of this section the smearing procedure produces a spherical cavity of radius $\tilde r_q$. Inside this cavity the solution is sourced solely by the colour D3--branes through the self dual RR $F_{5}$ form. Before presenting the solution for the functions of the first order perturbative expansion \eqref{expansion1}, we will elaborate more on the equations of motion inside the cavity.

%%%%%%%%%%%%%%%%%%%%%%%%%%%%%%%%%%%%%%%%%%%%%%%%%%%%%%%%%%%%%%%%%%%%%%%%%%%%%%%%%%%%

\subsection{Vacuum solution inside the cavity}

The smearing of massive flavour D7--branes produces a hollow spherical cavity at the origin of the subspace transverse to the colour D3--branes. 
Inside this cavity the gravitational background is a solution of the vacuum equations of motion (setting $Q_f = 0$ in  equations \eqref{defJ} \& \eqref{diff-b}--\eqref{diff-H}). 
It turns out that it is possible to obtain non-perturbative solutions for all background functions, except for the functions $S$ and $F$ describing the radii  of the $CP_2$  internal subspace and corresponding $U(1)$ fiber bundle. %It is the scope of this section to present the solutions.

%%%%%%%%%%%%%%%%%%%%%%%%%%%%%%%%%%%%%%%%%%%%%%%%%%%%%%%%%%%%%%%%%%

\subsubsection{The equations of motion in the vacuum}

The equations of motion inside the cavity are given by 
\begin{eqnarray}
\partial_\sigma^2(\log b)&=&-\frac{e^{-\Phi}\,h\,H'^2}{b^2} -e^{\Phi}\,h\,J'^2 \, ,
\label{diff-bv}
\\ \nonumber \\
\partial_\sigma^2(\log h)&=&-Q_c^2\frac{b^2}{h^2}-\frac{1}{2}\frac{e^{-\Phi}\,h\,H'^2}{b^2} -\frac{3}{2}e^{\Phi}\,h\,J'^2  \, ,
\label{diff-hv}
\\ \nonumber \\
\partial_\sigma^2(\log S)&=& -2\, b^2 F^4 S^4 + 6\, b^2 F^2 S^6+\frac{1}{4}\frac{e^{-\Phi}\,h\,H'^2}{b^2} +\frac{1}{4}e^{\Phi}\,h\,J'^2   \, ,
\label{diff-Sv}
\\ \nonumber \\
\partial_\sigma^2(\log F)&=& 4\,b^2 F^4 S^4+\frac{1}{4}\frac{e^{-\Phi}\,h\,H'^2}{b^2} +\frac{1}{4}e^{\Phi}\,h\,J'^2  \, ,
\label{diff-Fv}
\\ \nonumber \\
\partial_\sigma^2\Phi&=& -\frac{1}{2}\frac{e^{-\Phi}\,h\,H'^2}{b^2} -\frac{1}{2}e^{\Phi}\,h\,J'^2   \, ,
\label{diff-Phiv}
\\ \nonumber \\
\partial_{\sigma} \left[\frac{e^{-\Phi}\,h\,H'}{b^2}\right]&=& \,Q_c\,J'\, ,
\label{diff-Hv}
\\ \nonumber \\
\partial_{\sigma} \left[e^{\Phi}\,h\,J'\right]&=& \,Q_c\,H'\, .
\label{diff-Jv}
\end{eqnarray}
Adding and subtracting in various ways \eqref{diff-bv}, \eqref{diff-Sv},  \eqref{diff-Fv} \& \eqref{diff-Phiv} 
we easily obtain the following system of equations without sources 
\begin{eqnarray}
\partial_\sigma^2\log \frac{F^2}{S^2} \, &=&  \, - \, 12\, \left(b \, S^4\right)^2 \, \frac{F^2}{S^2} \, \left(1 \, - \, \frac{F^2}{S^2}\right) \, ,
\label{diff-F/S}
\\ \nonumber \\
\partial_\sigma^2\log(bS^4) \, &=& \,  8\, \left(b \, S^4\right)^2 \, \frac{F^2}{S^2} \, \left(3 \, - \, \frac{F^2}{S^2}\right) \, , 
\label{diff-bS^4}
\\ \nonumber \\ 
\partial_\sigma^2\log e^{-2\Phi}b \, &=& \, 0 \label{diff-be^-2Phi}\ .
\end{eqnarray}
While \eqref{diff-be^-2Phi} strongly suggests $e^{2\Phi}\propto b$ for \eqref{diff-F/S} \& \eqref{diff-bS^4}  we need to define a new set of variables
\begin{equation}
U \, \equiv \, b \, S^4 \qquad \& \qquad  V \, \equiv \,\frac{F^2}{S^2} \, .
\end{equation}
In these variables the equations of motion for $V$ \& $U$ decouple in the following way
\begin{eqnarray}
\partial_\sigma^2\log U \, & = & \, 8U^2V \, \left(3 \, - \, V \right) \, , 
\label{diff-U}
\\ \nonumber \\
\partial_\sigma^2\log V \, & = & \, - \, 12 U^2V \, \left( 1 \, - \, V \right)\ .
\label{diff-V} 
\end{eqnarray}
Notice that $V$ is the ratio between the radii of the $CP_2$ and the fiber in the $S^5$,  therefore it is a measure for the relative squashing. 
The following solution to \eqref{diff-U} \& \eqref{diff-V}
\begin{equation}
V(\sigma) \, = \,  1 \quad \& \quad  U(\sigma) \, = \, \frac{1}{4\sigma+{\rm const}}\, \ , \label{SUSY-S5}
\end{equation} 
corresponds to a non-squashed $S^5$. Since in our case the vacuum solution at the boundary of the cavity should match the flavour background,
we are interested in deformations of \eqref{SUSY-S5} corresponding to a squashed $S^5$. We have no reason to 
expect enhancement of the global symmetry of the theory in the deep IR. 
From a field theory point of view the squashing of the $S^5$ corresponds to a breaking of the global $SU(4)$ symmetry down to $SU(3)\times U(1)$. Inside the cavity the gravitational background corresponds to the effective field theory obtained after integrating out the flavours. 

Now let us briefly discuss the rest of the vacuum equations of motion. A natural candidate for a solution consistent with our ansatz is the supergravity background dual to a non-commutative Yang-Mills, which can be obtained as a near horizon limit of the supergravity solution corresponding to the D3-D1 bound state. 
Indeed one can check that upon the following substitution
\begin{eqnarray}
b(\hat r) \, &=& \, \frac{1 +  c_b\Theta^2}{1+\Theta^4\hat r^4} \, , 
\label{sol-b} \\ \nonumber \\
e^{2\Phi(\hat r)} \, &=& \, e^{2\Phi_*}\frac{1+c_{\Phi}\Theta^2}{1+\Theta^4\hat r^4}\, , 
\\ \nonumber \\
H(\hat r) \, &=& \, H_{*}-\Theta^2e^{\frac{\Phi_*}{2}}\frac{\hat r^4}{R^2} 
\frac{(1+c_b\Theta^2)^{\frac{1}{2}}}{1+\Theta^4\hat r^4}(1+c_{\Phi}\Theta^2)^{\frac{1}{4}} \, , 
\label{sol-H}\\ \nonumber \\
J(\hat r) \, &=& \, \Theta^2e^{-\frac{\Phi_*}{2}}\frac{\hat r^4}{R^2}(1+c_b\Theta^2)^{-\frac{1}{2}}(1+c_{\Phi}\Theta^2)^{-\frac{1}{4}} \, , 
\\ \nonumber \\
h(\hat r) \, &=& \, \frac{R^4}{\hat r^4}\frac{1+c_b\Theta^2}{\sqrt{1+\Theta^4\hat r^4}} \, ,
\label{sol-h}\\ \nonumber \\
\sigma(\hat r)&=&\frac{1}{4\hat r^4}+c_\sigma\Theta^2 \, ;~~~~\Theta^2\equiv e^{\frac{\Phi_*}{2}}\Theta^{23}/R^2\ , 
\label{sol-sigm}
\end{eqnarray} 
equations \eqref{diff-bv}, \eqref{diff-hv}, \eqref{diff-Phiv}, \eqref{diff-Hv} \& \eqref{diff-Jv} are satisfied where $\Theta^{23}$ is the parameter of non-commutativity of the dual field theory in the ($x^2,x^3$)-plane. 
The constants of integration $c_b\, ,c_{\Phi}$ and $c_\sigma$ will be fixed by matching to the flavour part of the background at the boundary of the cavity.  

Notice that the system of equations \eqref{diff-U} \& \eqref{diff-V} is completely decoupled and describes the different deformations of the compact part of the geometry. 
Expressions \eqref{sol-b}-\eqref{sol-sigm} together with \eqref{SUSY-S5} constitute a full solution  corresponding to the gravity dual of a  non-commutative SYM 
with $SU(4)$ global symmetry. It is possible to find a more general class of solutions corresponding to a non-commutative SYM with 
$SU(3)\times U(1)$ global  symmetry and this will be the topic of the next subsection.

%%%%%%%%%%%%%%%%%%%%%%%%%%%%%%%%%%%%%%%%%%%%%%%%%%%%%%%%%%%%%%%%%%%%%%%%%%%%

\subsubsection{Deforming the sphere} 

We now construct a non-commutative non-supersymmetric background by considering perturbative solutions of the system \eqref{diff-U} \& \eqref{diff-V} around \eqref{SUSY-S5} starting from the following ansatz:
\begin{equation}
U(\sigma)=\frac{1}{4\sigma} \, \left(1+\Sigma\,u_1 \right) \quad \& \quad V(\sigma) \, = \, 1+\Sigma\,v_1  \,  ,
\label{pert-UV}
\end{equation} 
where $u_1$ \& $v_1$ are functions of $\sigma$ to be determined after substituting in \eqref{diff-U} \& \eqref{diff-V} and expanding in $\Sigma$. 
At first order we have 
\begin{equation} 
u_1''-\frac{2}{\sigma^2} \, u_1-\frac{1}{2\sigma^2} \, v_1 \, = \, 0 
\quad \& \quad 
v_1'' - \frac{3}{4\sigma^2} \, v_1 \, = \, 0 \, ,   
\label{FOSigma}
\end{equation}
and after an appropriate redefinition 
\begin{equation}
\beta \, \equiv \, \frac{5}{2} \, u_1 +v_1  
\quad \Rightarrow \quad 
\beta''-\frac{2}{\sigma^2} \, \beta \, = \, 0 
\quad \& \quad
 v_1''-\frac{3}{4\sigma^2} \, v_1 \, = \, 0  \, .
\end{equation}
The general solution is given by 
\begin{equation}
\beta(\sigma) \, = \, \tilde c_{1}\sigma^2 + \tilde c_{2}\sigma^{-1} 
\quad \& \quad
 v_1 \, = \, \tilde c_3\sigma^{3/2} + \tilde c_4\sigma^{-1/2}  \,  ,
\end{equation}
and in terms of the radial variable $\hat r$ defined in \eqref{sol-sigm} the  functions $u_1$  \& $v_1$ are given by
\begin{eqnarray}
u_1&=&c_1\frac{1}{\hat r^8}+c_2{\hat r^4}-\frac{2}{5}c_3\hat r^{-6}-\frac{2}{5}c_4\hat r^2\ , \\
v_1&=&c_3\hat r^{-6}+c_4\hat r^2\ , 
\label{pert-uv}
\end{eqnarray}
where the constants $c_i$ are related to the constants $\tilde c_i$  by a non-singular diagonal linear transformation. 
Notice that in the formalism of \eqref{pert-UV} the parameter of non-commutativity $\Theta^2$ is proportional to $\Sigma$ and hence  $\sigma=1/(4\hat r^4)+O(\Sigma)$. 
Analyzing \eqref{pert-uv}, the general solution regular at $\hat r=0$ is spanned by $c_2\neq0, c_4\neq0\ , c_1=c_3=0$. The constants $c_2$ and $c_4$ will be determined by matching to the flavour solution at the boundary of the cavity (after setting $\Sigma \sim \epsilon_*$).

%%%%%%%%%%%%%%%%%%%%%%%%%%%%%%%%%%%%%%%%%%%%%%%%%%%%%%%%%%%%%%%%%%%%%%

\subsection{Constructing the full perturbative solution}

Our construction is a perturbative expansion near a supersymmetric AdS$_5\times S^5$ background. Requiring that the leading order corrections inside and outside the cavity agree implies that $\Theta^2\sim\epsilon_*$. 
Furthermore, we introduce the radial variable $r$ related to $\sigma$ via $r^4\equiv\frac{1}{4\sigma}$ and its dimensionless analog $\tilde r$ defined in equation (\ref{dimless-r}). 
The resulting expansions are
\begin{eqnarray}
b&=&1+c_b\Theta^2 \, ,
\label{per-b}\\ 
\Phi&=&\Phi_*+\frac{1}{2} \, c_{\phi} \, \Theta^2 \, , 
\label{per-Phi}\\
H(r)&=&H_*-\Theta^2e^{\frac{\Phi_*}{2}}\,\frac{r_m^4}{R^2}\,\tilde r^4 \, ,
\label{per-H}\\
J(r)&=&\Theta^2e^{-\frac{\Phi_*}{2}}\,\frac{r_m^4}{R^2} \, {\tilde r^4} \,  ,
\label{per-J}\\
h(r)&=&\frac{R^4}{r^4}\Bigg[1+\Theta^2\left(c_b-4c_{\sigma}r_m^4\tilde r^4\right)\Bigg]\ .
\label{per-h}
\end{eqnarray}
Since \eqref{per-H} does not depend on any constants of integration we can use the perturbative expansion of the flavour solution to determine 
the exact relation between $\Theta^2$ and $\epsilon_*$. Indeed comparing \eqref{per-H} and \eqref{expansion1}
 we conclude that for $\tilde r\leq \tilde r_q$ we have 
\begin{equation}
\epsilon_*H_1(\tilde r)=-\Theta^2e^{\frac{\Phi_*}{2}}\,\frac{r_m^4}{R^2}\,\tilde r^4\, ,  \label{H1-1}
\end{equation}
where $\tilde r_q$ is the radius of the cavity.
In the following subsections we will present semi-analytic solutions for all the functions of the first order perturbative expansion,  
requiring specific behaviors inside the cavity emerging from the expansions \eqref{per-b}-\eqref{per-h}.

%%%%%%%%%%%%%%%%%%%%%%%%%%%%%%%%%%%%%%%%%%%%%%%%%%%%%%%%%%%%%%%%%%%%%%%%%%%%%%%%%%%

\subsubsection{Solving for the $B$-field}

In order to solve the equation of motion for $H_1$ in \eqref{EOMs} we need the Greens function $G_{H_1}(\tilde r, \tilde r_1)$ satisfying the following equation
\begin{eqnarray}
\frac{d^2}{d{\tilde r}^2} \, G(\tilde r,\tilde r_1) \, + \, 
\frac{1}{\tilde r} \, \frac{d}{d\tilde r} \, G(\tilde r,\tilde r_1) \, - \, 
\frac{16}{\tilde r^2} \, G(\tilde r,\tilde r_1)\,
 & = & \, 
\delta(\tilde r-\tilde r_1) 
\nonumber \\ \nonumber \\
\quad  \quad \quad  \Rightarrow \quad  \quad \quad
G_{H_1}(\tilde r,\tilde r_1) \, &  = & \, \frac{1}{8} \, \frac{{\tilde r}^4}{{\tilde r_1^3}} \, \left(1\, - \, \frac{\tilde r_1^8}{\tilde r^8} \right) \,  
\Theta(\tilde r-\tilde r_1) \, .
\label{Green-H1}
\end{eqnarray}
The solution for $H_1$ that inside the cavity is $\propto\tilde r^4$ and vanishes at $\tilde r=\tilde r_*$ 
can be written in the following integral form
\begin{equation} \label{semi-H1}
H_1(\tilde r) \, = \, \frac{1}{2} \, \int\limits_{\tilde r_q}^{\tilde r_*}d\tilde r_1 \, J_{H_1}({\tilde r}_1)
\Bigg[G_{H_1}(\tilde r,\tilde r_1) \, - \, \frac{\tilde r^4}{\tilde r_*^4} \, G_{H_1}(\tilde r_*,\tilde r_1)\Bigg] \, .
\end{equation}
Elaborating on \eqref{semi-H1} inside the cavity ($\tilde r\leq\tilde r_q$) we have
\begin{equation}
H_1(\tilde r)=-H_*\#\left(\tilde r_q,\tilde r_*\right)
\,\tilde r^4\propto\tilde r^4 \, ,
\label{H1-2}
\end{equation}
where the expression for the constant $\#\left(\tilde r_q,\tilde r_*\right)$  is
\begin{equation} \label{constant}
\#\left(\tilde r_q,\tilde r_*\right)=\int\limits_{\tilde r_q}^{\tilde r_*}d\tilde r\frac{\left(1-\frac{\tilde r^8}{\tilde r_*^8}\right) }{2\tilde r^3\sqrt{{\tilde r}^4 \, + 1}}
\tilde\xi\left(\tilde r,\tilde r_q\right) \, .
\end{equation}
Comparing \eqref{H1-1} and \eqref{H1-2} we obtain the promised relation connecting the non-commutativity parameter to the number of flavours   
\begin{equation}
\Theta^2=\#\left(\tilde r_q,\tilde r_*\right)\frac{1}{r_m^2}\, \epsilon_*\propto \frac{N_f}{N_c}\frac{1}{H_*}\,  .
\label{non-com}
\end{equation}
%
%We postpone the detailed study of dependence of the parameter of non-commutativity on the physical parameters of the theory for one of the next subsections.

%%%%%%%%%%%%%%%%%%%%%%%%%%%%%%%%%%%%%%%%%%%%%%%%%%%%%%%%%%%%%%%%%%%%%%%%%%%

\subsubsection{Solving for the dilaton \&  $b_1$}

In order to solve the equations of motion for both the dilaton and $b_1$ in \eqref{EOMs} we need the Greens function 
$G_{\{\Phi_1, b_1\}}(\tilde r, \tilde r_1)$ satisfying the following equation
\begin{eqnarray}
\frac{d^2}{d{\tilde r}^2} \, G(\tilde r,\tilde r_1) \, + \, 
\frac{5}{\tilde r} \, \frac{d}{d\tilde r} \, G(\tilde r,\tilde r_1) \,
 & = & \, 
\delta(\tilde r-\tilde r_1) 
\nonumber \\ \nonumber \\
\quad  \quad \quad  \Rightarrow \quad  \quad \quad
G_{\{\Phi_1, b_1\}}(\tilde r,\tilde r_1) \, &  = & \,\frac{\tilde r_1}{4} \, \left(1\, - \, \frac{\tilde r_1^4}{\tilde r^4} \right) \,  
\Theta(\tilde r-\tilde r_1) \, .
\label{Green-H1}
\end{eqnarray}
Looking at  \eqref{per-b} and \eqref{per-Phi} we see that both the dilaton and $b$ are constant inside the cavity. 
The solution for $\Phi_1$ \& $b_1$ which is constant inside the cavity and vanishing at $\tilde r=\tilde r_*$ 
can be written in the following integral form
\begin{eqnarray}
\Phi_1(\tilde r) \, &=& \, 
\frac{1}{2} \, \int\limits_{\tilde r_q}^{\tilde r_*}d\tilde r_1 \, J_{\Phi_1}(\tilde r_1) \, 
\Bigg[G_{\{\Phi_1, b_1\}}(\tilde r,\tilde r_1) \, - \, G_{\{\Phi_1, b_1\}}(\tilde r_*,\tilde r_1) \Bigg] \, ,
\label{semi-Phi1} \\
b_1(\tilde r) \, &=& \, 
\frac{1}{2} \, \int\limits_{\tilde r_q}^{\tilde r_*}d\tilde r_1 \, J_{b_1}(\tilde r_1) \, 
\Bigg[G_{\{\Phi_1, b_1\}}(\tilde r,\tilde r_1) \, - \, G_{\{\Phi_1, b_1\}}(\tilde r_*,\tilde r_1)\Bigg] \, .
\label{semi-b1}
\end{eqnarray}
Inside the cavity both $\Phi_1$ and $b_1$ are constant and are given by the following expressions
\begin{eqnarray}
\Phi_1(\tilde r)&=& \, - \, \frac{1}{2} \, \int\limits_{\tilde r_q}^{\tilde r_*}d\tilde r_1 \, J_{\Phi_1}(\tilde r_1)\, G_{\{\Phi_1, b_1\}}(\tilde r_*,\tilde r_1) \, ,
\label{phi1-const}\\
b_1(\tilde r)&=& \, - \, \frac{1}{2} \, \int\limits_{\tilde r_q}^{\tilde r_*}d\tilde r_1 \, J_{b_1}(\tilde r_1)\, G_{\{\Phi_1, b_1\}}(\tilde r_*,\tilde r_1)\, \ .\label{b1-const}
\end{eqnarray}
Equations \eqref{phi1-const}, \eqref{b1-const} and \eqref{non-com} can be used to fix the constants $c_b$ and $c_{\Phi}$.

%%%%%%%%%%%%%%%%%%%%%%%%%%%%%%%%%%%%%%%%%%%%%%%%%%%%%%%%%%%%%%%

\subsubsection{Solving for $\lambda_1$}

Proceeding in the same way we obtain the Green's function for the equation of motion for $\lambda_1$, defined in \eqref{EOMs} :
\begin{eqnarray}
\frac{d^2}{d{\tilde r}^2} \, G(\tilde r,\tilde r_1) \, + \, 
\frac{5}{\tilde r}\frac{d}{d\tilde r} \, G(\tilde r,\tilde r_1) \, - \, 
\frac{32}{\tilde{r}^2} \, G(\tilde r,\tilde r_1) \, 
& = &  
\, \delta(\tilde r-\tilde r_1)
\nonumber \\ \nonumber \\
\quad  \quad \quad  \Rightarrow \quad  \quad \quad
G_{\lambda_1}(\tilde r,\tilde r_1) \, 
& = & 
\, \frac{\tilde r^4}{12 \, \tilde r_1^3} \, 
\left(1\, - \, \frac{\tilde r_1^{12}}{\tilde r^{12}} \right) \,  \Theta(\tilde r-\tilde r_1) \, ,
\label{Green-lambda1}
\end{eqnarray}
and look for a solution that inside the cavity is $\propto \tilde r^4$ and vanishes at $\tilde r=\tilde r_*$
\begin{equation}
\lambda_1 \, = \, \frac{1}{2} \, 
\int\limits_{\tilde r_q}^{\tilde r_*}d\tilde r_1 \, J_{\lambda_1}({\tilde r}_1)   \, 
\Bigg[G_{\lambda_1}(\tilde r,\tilde r_1) \, - \, \frac{\tilde r^4}{\tilde r_*^4} \, G_{\lambda_1}(\tilde r_*,\tilde r_1)\Bigg]\ .
\label{semi-lambda1}
\end{equation}
%

%%%%%%%%%%%%%%%%%%%%%%%%%%%%%%%%%%%%%%%%%%%%%%%%%%%%%%%%%%%%%%%%%%

%%%%%%%%%%%%%%%%%%%%%%%%%%%%%%%%%%%%%%%%%%%%%%%%%%%%%%%%%%%%%%%%%%%%%%%

\subsubsection{Solving for $\Delta_1$}

Proceeding in the same way for $\Delta_1$ we have 
\begin{eqnarray}
\frac{d^2}{d{\tilde r}^2} \, G(\tilde r,\tilde r_1) \, + \, 
\frac{5}{\tilde r}\frac{d}{d\tilde r} \, G(\tilde r,\tilde r_1) \, - \, 
\frac{12}{\tilde{r}^2} \, G(\tilde r,\tilde r_1) \, 
& = &  
\, \delta(\tilde r-\tilde r_1)
\nonumber \\ \nonumber \\
\quad  \quad \quad  \Rightarrow \quad  \quad \quad
G_{\Delta_1}(\tilde r,\tilde r_1) \, 
& = & 
\, \frac{\tilde r^2}{8 \, \tilde r_1} \, 
\left(1\, - \, \frac{\tilde r_1^{8}}{\tilde r^{8}} \right) \,  \Theta(\tilde r-\tilde r_1) \, .
\end{eqnarray}
Looking for a solution which behaves as  $\Delta_1(\tilde r)\propto \tilde r^2$ inside the cavity and asymptotes to the supersymmetric solution at $\tilde r_*$
\begin{equation}
\eqref{susy-Delta1} \qquad \Rightarrow \qquad  \Delta_1({\tilde r}_*) \, = \, - \, 
\frac{1}{12} \, \left(1 \, - \frac{{\tilde m}_0^2}{{\tilde r}_*^2} \right)^3\ ,
\end{equation}
we have 
\begin{equation}
\Delta_1(\tilde r) \,  = \, \frac{1}{2} \, 
 \int\limits_{\tilde r_q}^{\tilde r_*}d\tilde r_1 \, J_{\Delta_1}(\tilde r_1) \,  
\Bigg[G_{\Delta_1}(\tilde r,\tilde r_1)-\frac{\tilde r^2}{\tilde r_*^2}G_{\Delta_1}(\tilde r_*,\tilde r_1)\Bigg] \, - \, 
\frac{1}{12} \, \frac{\tilde r^2}{\tilde r_*^2} \, \left(1 \, - \frac{{\tilde m}_0^2}{{\tilde r}_*^2} \right)^3\, .
\label{semi-Delta1}
\end{equation}

%%%%%%%%%%%%%%%%%%%%%%%%%%%%%%%%%%%%%%%%%%%%%%%%%%%%%%%%%%%%%%%%%%%%%%%

\subsubsection{Solving for $\Lambda_1$}

The Greens function for $\Lambda_1$ is given by \eqref{Green-lambda1}  so we proceed immediately in looking for a solution 
which behaves as  $\Lambda_1(\tilde r)\propto \tilde r^4$ inside the cavity and asymptotes to the supersymmetric solution at $\tilde r_*$
\begin{equation}
\eqref{susy-Lambda1} \qquad \Rightarrow \qquad  \Lambda_1({\tilde r}_*) \, = \, \ ,
\frac{1}{18} \, \left(1 \, - \frac{{\tilde m}_0^2}{{\tilde r}_*^2} \right)^3
\end{equation}
we have 
\begin{equation}
\Lambda_1(\tilde r) \,  = \, \frac{1}{2} \, 
 \int\limits_{\tilde r_q}^{\tilde r_*}d\tilde r_1 \, J_{\Lambda_1}(\tilde r_1) \,  
\Bigg[G_{\Lambda_1}(\tilde r,\tilde r_1)-\frac{\tilde r^4}{\tilde r_*^4}G_{\Lambda_1}(\tilde r_*,\tilde r_1)\Bigg] \, + \, 
\frac{1}{18} \, \frac{\tilde r^4}{\tilde r_*^4} \, \left(1 \, - \frac{{\tilde m}_0^2}{{\tilde r}_*^2} \right)^3\, .
\label{semi-Lambda1}
\end{equation}

%%%%%%%%%%%%%%%%%%%%%%%%%%%%%%%%%%%%%%%%%%%%%%%%%%%

\subsubsection{Solving the constraint for $\eta_1$}

The first order differential equation for $\eta$ \eqref{EOM-eta} can be integrated to give
\begin{equation}
\eta_1({\tilde r}) \, = \, \tilde r^4 \Bigg[ \frac{\eta_1({\tilde r}_*)}{{\tilde r}_*^4} \, 
+ \, \int \limits_{{\tilde r}}^{\tilde r_*}d\hat r \, \frac{J_{\eta_1}({\hat r})}{\hat r^4}\,\Theta(\tilde r-\tilde r_q)  \Bigg]
\quad \text{with} \quad 
\eta_1(\tilde r_*) \, \Rightarrow \, \frac{1}{9}  \, \left(1 \, - \frac{{\tilde m}_0^2}{{\tilde r}_*^2} \right)^3 \, . 
\end{equation}
The value of  $\eta_1(\tilde r_*)$  is fixed by the requirement of obtaining the supersymmetric solution at  $\tilde r=\tilde r_*$.

\begin{figure}[h] %  figure placement: here, top, bottom, or page
   \centering
   \includegraphics[width=3.4in]{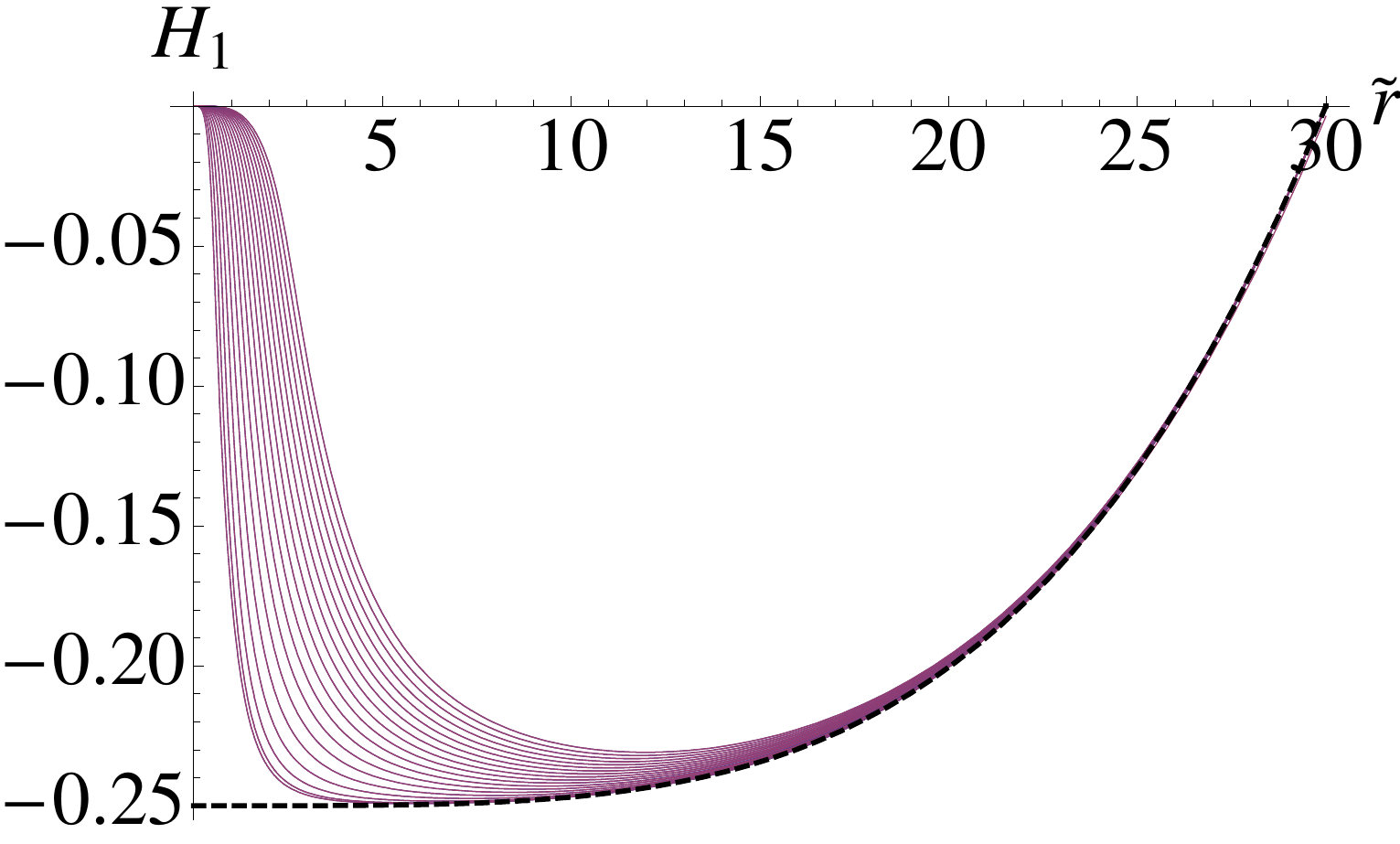} 
   \includegraphics[width=3.4in]{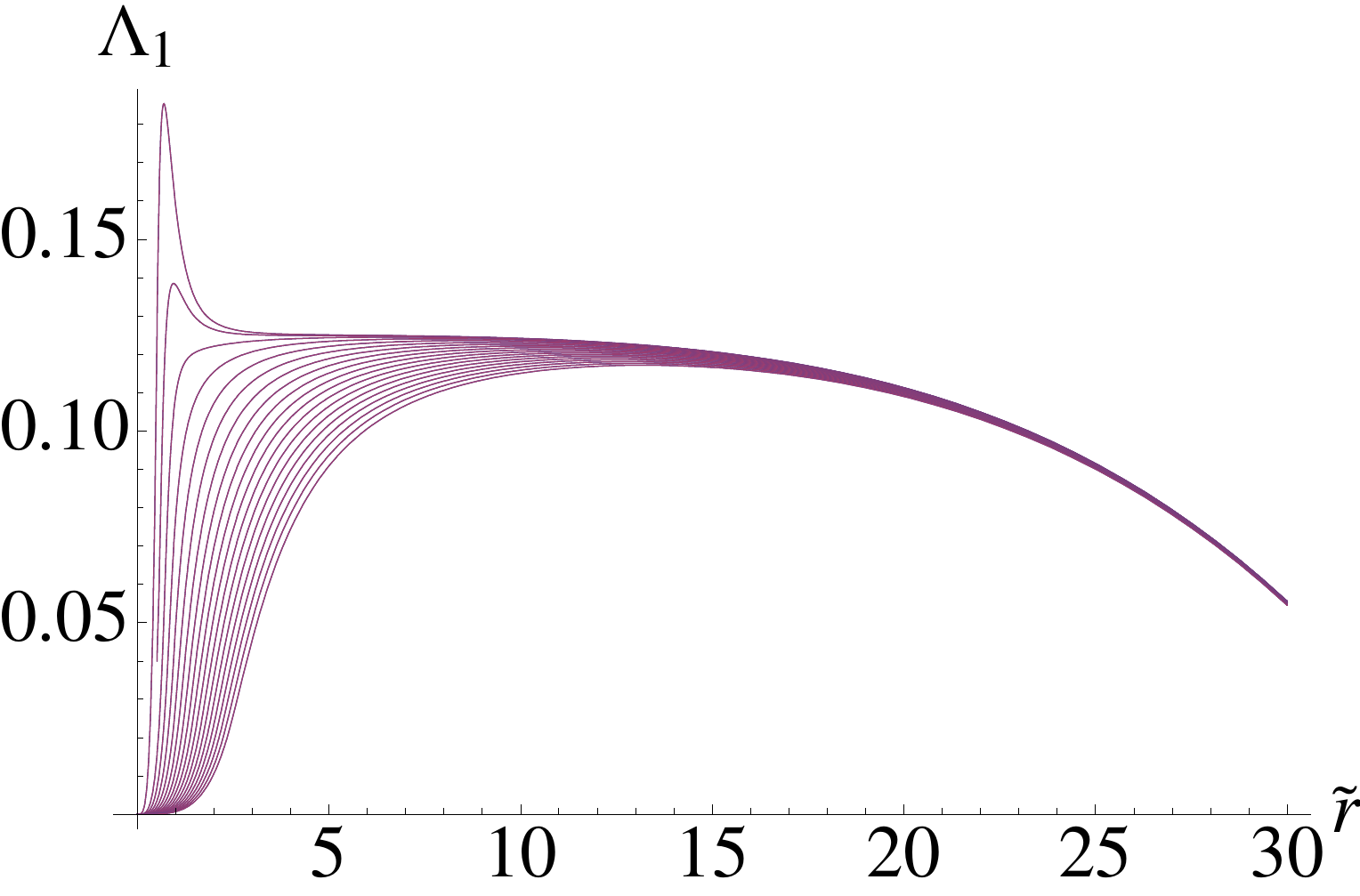} 
      \includegraphics[width=3.4in]{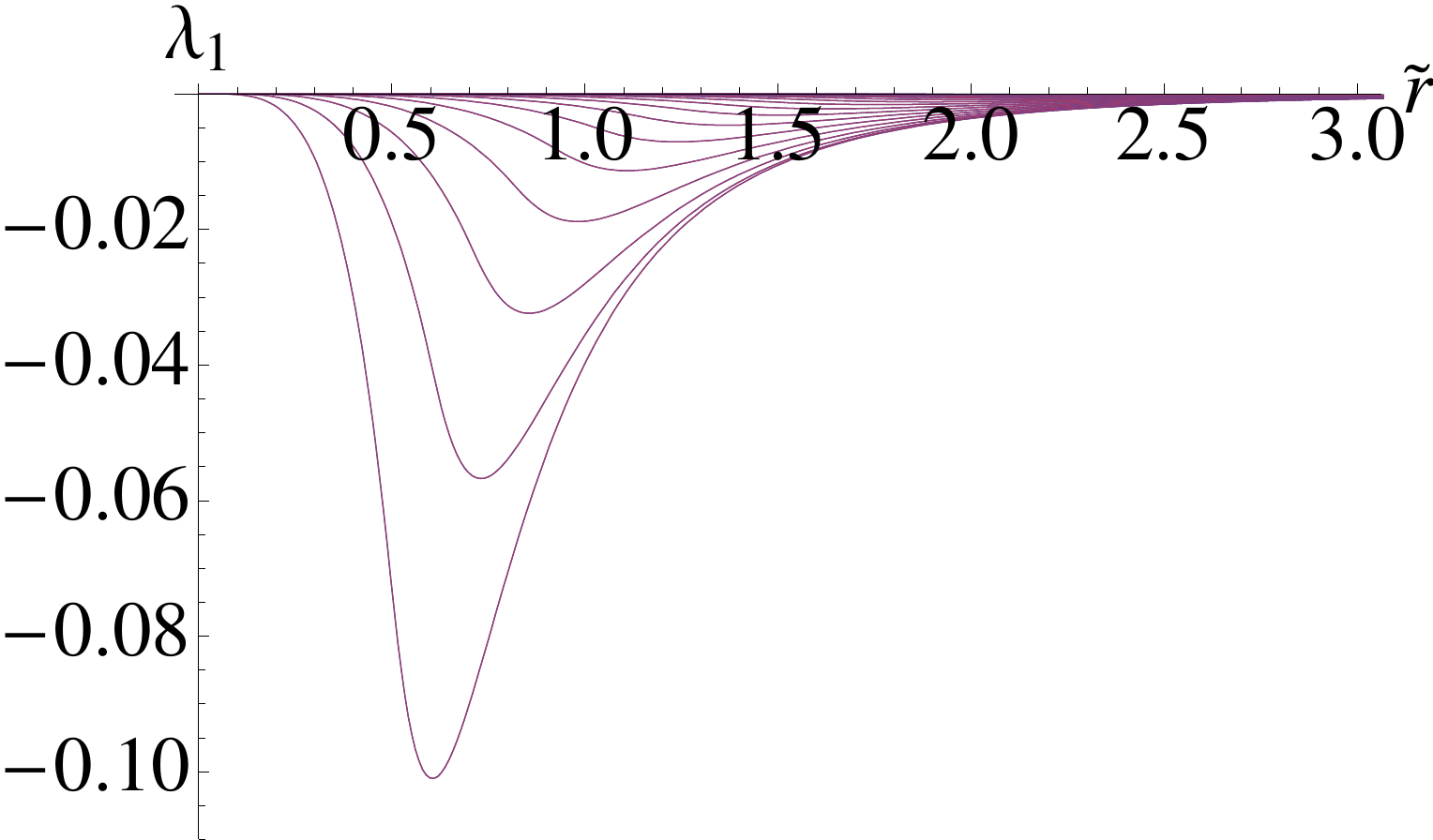} 
   \includegraphics[width=3.4in]{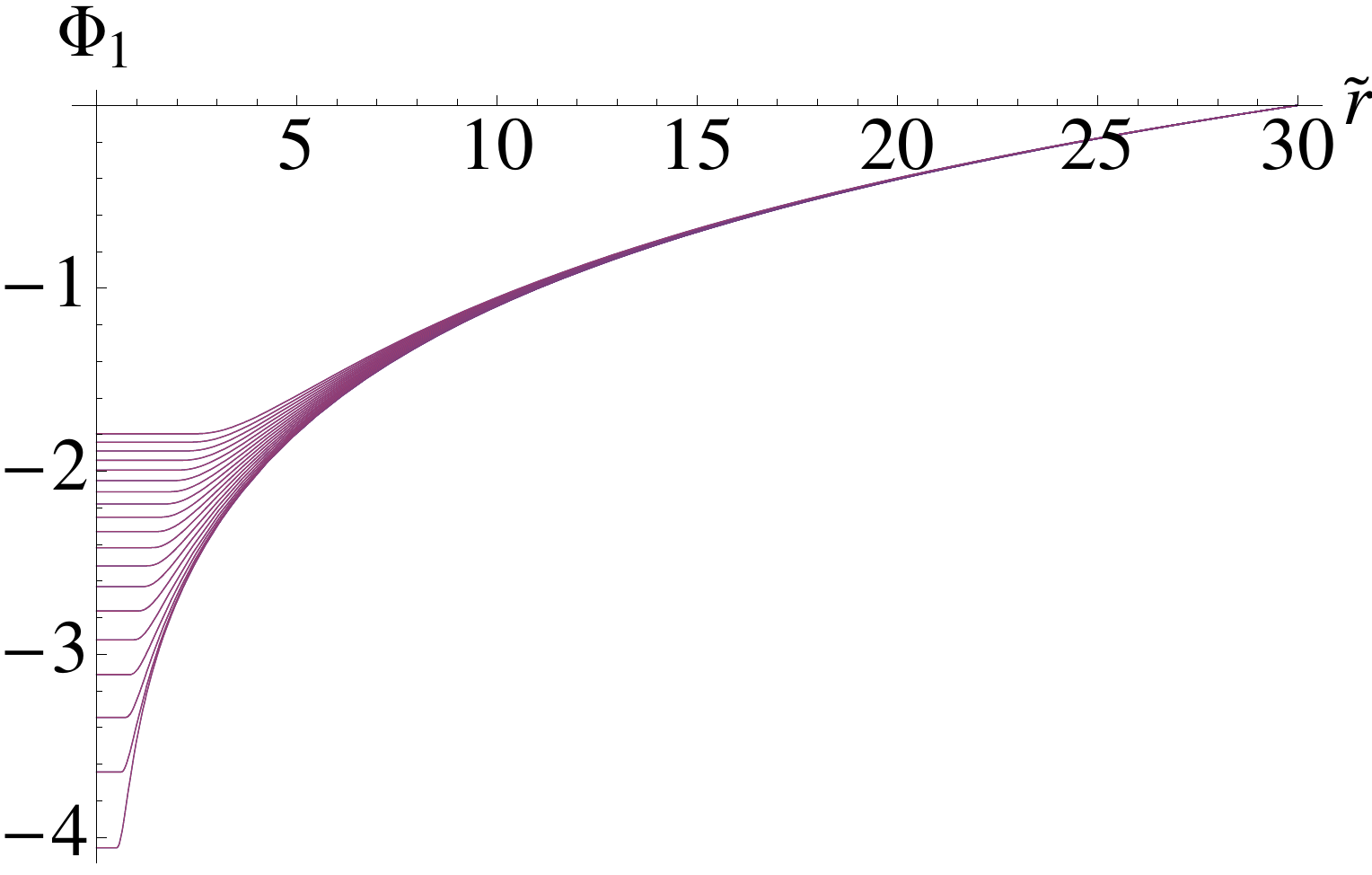} 
      \includegraphics[width=3.4in]{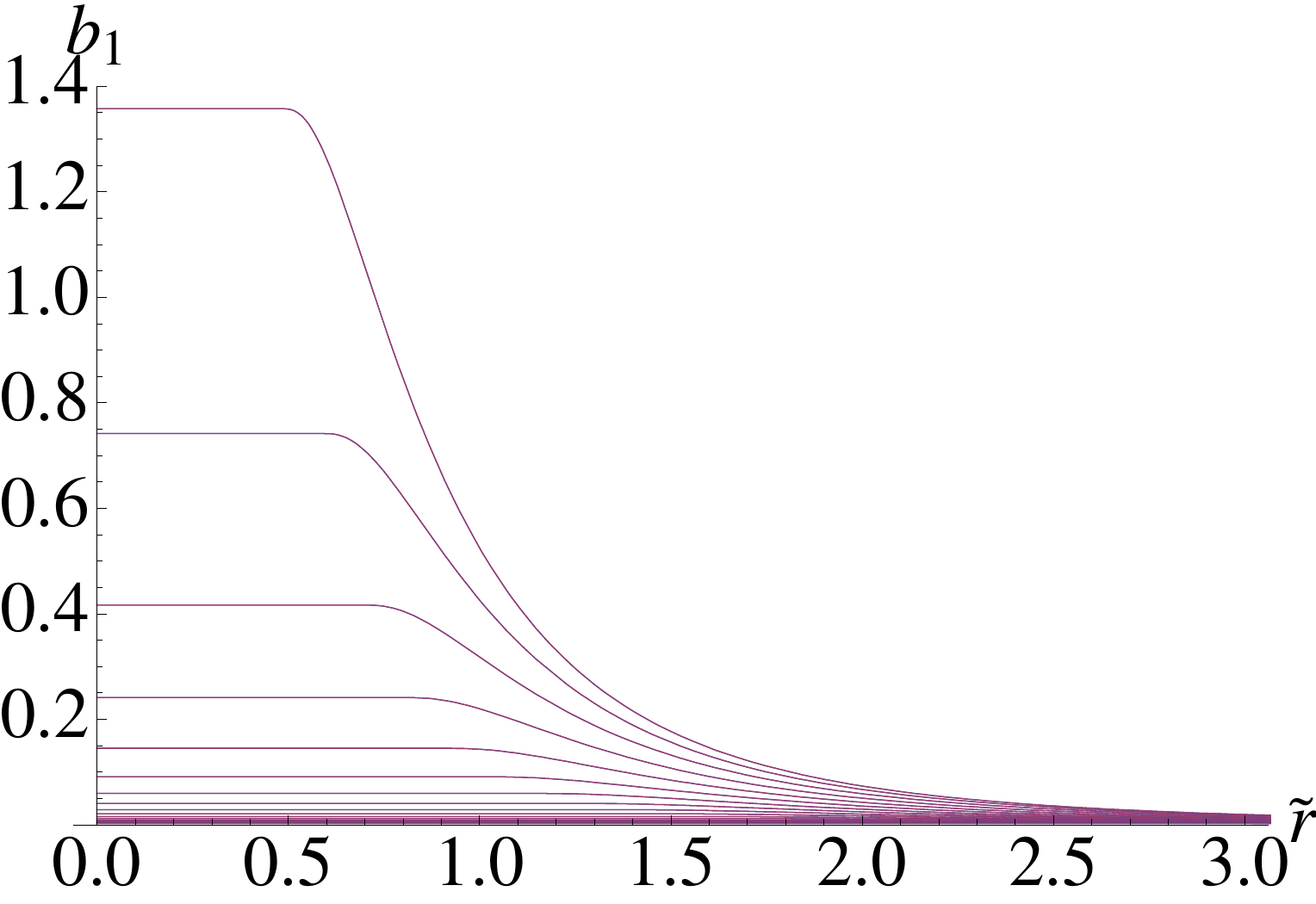} 
   \includegraphics[width=3.4in]{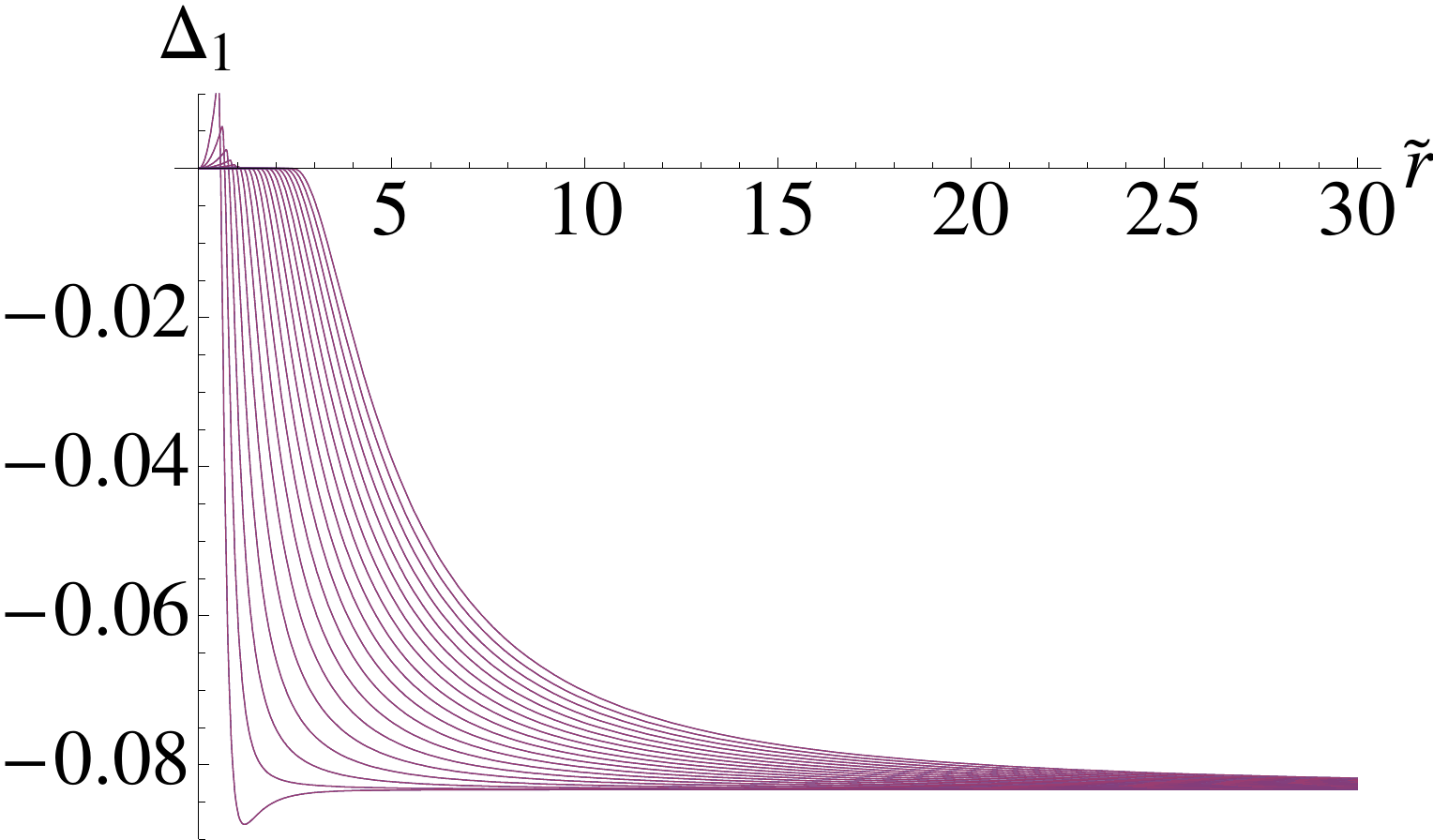} 
  \caption{Family of solutions of the background functions for different bare mass parameter $\tilde m$.}
   \label{fig:fig0}
\end{figure}
%

%%%%%%%%%%%%%%%%%%%%%%%%%%%%%%%%%%%%%%%%%%%%%%%%%%%

\subsection{Numerical solution}
As we learned from the previous subsection it is possible to obtain a solution for our supergravity background in a semi-analytic form in which all background functions are written in an integral form. The solution is completely determined by the D7--brane charge distribution function $p(\tilde r)$ or equivalently by the profile of the fiducial embedding $\chi_q(\tilde r)$. Furthermore, in the perturbative expansion the equations of motion for the first order corrections to the background fields are sourced by the zeroth order expansion of $\chi_q(\tilde r)$ (namely $\chi_0(\tilde r)$), which is known from the probe approximation. The equation of motion for $\chi_0(\tilde r)$ is non-linear and we solve it numerically.

An obvious approach to obtain a numerical solution for the background functions would be to solve numerically the integral expressions presented in section~2. Actually it is more convenient to solve the equations of motion completely numerically using Mathematica's built in function NDSolve. To generate the solution we employ standard shooting techniques and sew the numerical solution to the analytic solution inside the cavity at $\tilde r=\tilde r_q\,$. The constants of integration (which specify the shooting parameters) are obtained by matching the solution to the supersymmetric one at $\tilde r =\tilde r_*\,$. This approach has the disadvantage that the accumulated numerical error grows strongly with the parameter $\tilde r_*$, which we need to keep numerically high. It can be improved if we approximate the function $\chi_0$ with a taylor series expansion in $1/\tilde r$ for $\tilde r > \tilde r_{sew}$ ($\tilde r_{sew}$ is a number of order ten) and solve analytically for the background fields in the region $\tilde r_{sew}\leq \tilde r\leq \tilde r_*$ by matching the solution to the numerical one at $\tilde r_{sew}$ and to the supersymmetric one at $\tilde r_*$.

Repeating the procedure outlined above for different values of the radius of the cavity $\tilde r_q$ generates solutions corresponding to different quark masses. In figure \ref{fig:fig0} we have presented plots of the background functions for the following range of the bare mass parameter $\tilde m_0\in [0,2.3]$, which is the one needed for the analysis in section~5. As one can see the solutions corresponding to different values of $\tilde m$ differ significantly in the infrared $\tilde r \ll \tilde r_*$ and are very close to each other for $\tilde r\lesssim\tilde r_*$. In fact one can check that, except for the function $H_1$, for $\tilde r\lesssim \tilde r_*$ the background functions are very well approximated by their corresponding functions from the supersymmetric limit. The function $H_1$ follows the pattern $H_1(\tilde r)\approx\frac{1}{4}(\frac{\tilde r^4}{\tilde r_*}-1)$ represented by the black dashed curve in figure \ref{fig:fig0}. Apparently at $\tilde r_*$, $H_1(\tilde r_*)$ vanishes, but has a non-vanishing first derivative, therefore the matching to the supersymmetric solution is not smooth at $\tilde r_*$. However, one can check that $H_1'(\tilde r_*)=1/\tilde r_*$ which is sub-leading for $\tilde r_*\gg 1$. 

Overall we conclude that our solution is very well approximated by the supersymmetric background near $\tilde r_*$ and hence one can approximate the background with the supersymmetric one for $\tilde r >\tilde r_*$. This is particularly useful to investigate the UV behaviour of the background, because it relates non-perturbatively the arbitrary UV scale $r_*$ and the parameter of perturbative expansion $\epsilon_*$ to the position of the Landau pole $r_{\rm LP}$. In the next subsection we provide a detailed description of the hierarchy of scales and regime of validity of our perturbative solution.

%%%%%%%%%%%%%%%%%%%%%%%%%%%%%%%%%%%%%%%%%%%%%%%%%%%

\subsection{Hierarchy of scales and regime of validity }

In this subsection we analyze the regime of validity of our perturbative solution and the hierarchy of energy scales (in terms of radial scales) of the theory. Our analysis follows closely section 2.4 of ref.~\cite{arXiv:0909.2865}, where the finite temperature system has been unquenched. In terms of radial coordinates the hierarchy of scales can be written in the following way:
\begin{equation}
0<r_m\sim r_q \ll r_* \ll r_a < r_{\rm LP}\ ,
\end{equation}
where $r_m$ and $r_q$ represent IR energy scales related to the energy scale set by the magnetic field and the one corresponding to the physical mass of the fundamental matter. The radial scales $r_a$ and $r_{\rm LP}$ represent UV energy scales corresponding to the scale at which the supergravity solution develops pathologies and the Landau pole of the theory at which the effective 't Hooft coupling blows up. For energy scales close to $r_*$ our solution, as can be seen from the plots in figure {\ref{fig:fig0}}, is well approximated by the supersymmetric solution corresponding to vanishing magnetic field \cite{arXiv:0807.0298}, which (given our choice $r_q\ll r_*$) is well approximated by the supersymmetric solution corresponding to massless fundamental fields \cite{hep-th/0612118}. This is why the analysis of the UV energy scales is exactly the same as in ref.~\cite{arXiv:0909.2865} and in particular the relation between the ``position" of the Landau pole $r_{\rm LP}$, the finite cutoff $r_*$ and the perturbative parameter $\epsilon_*$ is:
\begin{equation}
\frac{r_*}{r_{\rm LP}}\approx e^{-1/\epsilon_*}\ll 1\ .
\end{equation}
Furthermore in order for our perturbative solution to be valid in the region $r_q\leq r\leq r_*$ we need $e^{\Phi(r_q)}/e^{\Phi_*}\sim 1$, which requires $\epsilon_*\,|\Phi_1(\tilde r_q)|\ll 1$ (because $|\Phi_1(\tilde r_q)|$ is the biggest off all first order corrections). However, as one can check from the plot in Figure \ref{fig:fig0}, ${\rm max}{}_{\tilde r_q}|\Phi_1(\tilde r_q)|\sim \log\frac{r_*}{r_q}$. Therefore we need $\frac{r_q}{r_*}\gg e^{-1/\epsilon_*}$.  

In addition we would like to be able to neglect terms sub-leading in the $r_q/r_*$ expansion. Therefore we have to make sure that the corrections $\epsilon_*$ that we are considering are much larger than the one that we ignore. This requires $\epsilon \gg \frac{r_q}{ r_*}$. In summary we have:
\begin{equation}
e^{-1/\epsilon_*}\ll\frac{M_q}{\Lambda_{\rm UV}}\sim \frac{r_q}{r_*}\ll\epsilon_*\sim \frac{\lambda_*\, N_f}{8\pi^2\, N_c}\ll 1\ .
\end{equation}
Finally, validity of the supergravity approximation requires that we ignore closed string loops ($N_c\gg 1$) and $\alpha'$ corrections ($\lambda_q \gg 1$), where $\lambda_q$ is the effective 't Hooft coupling at the energy scale set by $r_q\sim M_q$. It is related to $\lambda_*$ via $\lambda_q=\frac{e^{\Phi(r_q)}}{e^{\Phi_*}}\lambda_*$. In addition, validity of the smearing approximation suggests large $N_f$. In summary we have:
\begin{equation} 
N_c \gg 1,\quad \lambda_{q}\gg 1\ ,\quad \epsilon_{q}\equiv \frac{\lambda_q\,N_f}{8\pi^2\,N_c}\ll 1\ ,
\end{equation}
where we have defined the IR perturbative parameter $\epsilon_q$. Clearly it is related to $\epsilon_*$ (defined in equation \eqref{epsilonstar}) via:
\begin{equation}
\epsilon_{q}=\epsilon_*\frac{e^{\Phi(\tilde r_q)}}{e^{\Phi_*}}=\epsilon_*(1+\epsilon_*\Phi_1(\tilde r_q))+O(\epsilon_*^3)\ ,
\end{equation}
which also implies:
\begin{equation}
\epsilon_{*}=\epsilon_q(1-\epsilon_q\Phi_1(\tilde r_q))+O(\epsilon_q^3)\ .\label{epsstar}
\end{equation}
Finally requiring that $\alpha'$ corrections, which scale as $\lambda_q^{-3/2}$, (for more details look at ref.~\cite{arXiv:0909.2865}) are sub-leading relative to flavour corrections controlled by $\epsilon_q$ requires:
\begin{equation}
\lambda_q^{-3/2}\ll \epsilon_q \ .
\end{equation} 
We close this section by commenting that the numerical values used in our analysis are in the regime of validity specified above. 

%%%%%%%%%%%%%%%%%%%%%%%%%%%%%%%%%%%%%%%%%%%%%%%%%%%%%%%%%%%%%%%%%%%%%%%%%%%
%%%%%%%%%%%%%%%%%%%%%%%%%%%%%%%%%%%%%%%%%%%%%%%%%%%%%%%%%%%%%%%%%%%%%%%%%%%

\section{Free Energy and Condensate}

In this section we calculate the free energy density and fundamental condensate of the dual field theory directly from the supergravity background.

%%%%%%%%%%%%%%%%%%%%%%%%%%%%%%%%%%%%%%%%%%%%%%%%%%%%%%%%%%%%%%%%%%%%%%%%%%%

\subsection{Helmholtz versus Gibbs free energy}

Following the general prescription of \cite{Hawking:1995fd} we identify the on-shell Euclidean action with the Helmholtz free energy. 
The Euclidean action has contributions from two terms ${\cal I}_{\rm bulk}$ and ${\cal I}_{\rm surf}$ given by:
\begin{eqnarray} 
{\cal I}_{\rm bulk}&=&-\frac{V_{4}\pi^3}{2\kappa_{10}^2}\int {\cal L}_{\rm{II B}} \, d\sigma\ , \\
{\cal I}_{\rm surf}&=&-\frac{V_{4}\pi^3}{\kappa_{10}^2}\sqrt{\gamma}K \, \ ,\label{surf}
\end{eqnarray}
where ${\cal I}_{{\rm IIB}}$ is the wick rotated action (\ref{genact}) and in (\ref{surf}) is the standard Gibbons-Hawking term. 
In fact one can check that the one dimensional effective lagrangian ${\cal L}_{1d}$ defined in (\ref{L-effective1}) already includes this boundary term. 
Therefore we can write:
\begin{equation}
{\cal I}={\cal I}_{\rm bulk}+{\cal I}_{\rm surf}=-\frac{V_{4}\pi^3}{2\kappa_{10}^2}\int {\cal L}_{1d}(B,B',J',\psi_a,\psi_a') \, d\sigma\ ,
\end{equation}
where $\psi_a$ is a collective notation for the background functions $(b,h,F,S,\Phi)$. Note that:
\begin{equation}
\frac{\delta {\cal I}}{\delta J}\propto\frac{\partial {\cal L}_{1d}}{\partial J'}\propto H_*\ ,{\label{MagJ}}
\end{equation}
where we have used (\ref{defJ}). Given that we make the identification ${\cal I}/V_4={\cal F}/V_3$, where $\cal F$ refers to the Helmholtz free energy and we have that $\frac{\delta{\cal F}}{\delta M}=H_*$, it is natural to relate the field $J$ to the magnetization $M$ of the system. However we are interested in a thermodynamic ensemble in which we keep the external magnetic field $H_*$ fixed. Therefore the proper thermodynamic potential is given by the Gibbs free energy ${\cal G}\equiv{\cal F}-M H_*$. Equation (\ref{MagJ}) implies:
\begin{equation}
{\cal G} \, = \, \frac{V_3}{V_4} \, \tilde{\cal I} \, = \, - \, \frac{V_{3}\pi^3}{2\kappa_{10}^2}\int \tilde{\cal L}_{1d}(B,B',H_*,\psi_a,\psi_a') \, d\sigma\ ,\label{Gibbsfree}
\end{equation}
were $\tilde{\cal L}_{1d}$ refers to the Legendre transform of  ${\cal L}_{1d}$ defined in (\ref{L-effective2}). 
Next we expand in $\epsilon_*$ using (\ref{expansion1}) and define dimensionless variables along \eqref{dimless-r}. To first order one has:
\begin{equation}
\tilde{\cal I}={\cal I}_0+\epsilon_*{\cal I}_{\rm DBI}+\epsilon_*{\cal I}_{\rm bound}+O(\epsilon_*^2)\ ,
\end{equation}
and more explicitly:
\begin{eqnarray}
-\frac{2\kappa_{10}^2}{V_{3}\pi^3}{\cal G}&=&6r_*^4-4\epsilon_*r_m^4\int\limits_{0}^{\tilde r_*}d\tilde r\left[-\frac{d}{d\tilde r}\left(\tilde r^4\, \eta(\tilde r)\right)+\tilde r\sqrt{1+\tilde r^4}\tilde\xi(\tilde r)\right]={\label{Gibbs1}}\\
&=&\nonumber6r_*^4+4\epsilon_*r_*^4\eta(r_*)-4\epsilon_*\,r_m^4\,\int\limits_{\tilde r_q}^{\tilde r_*}d\tilde r\tilde r\,\sqrt{1+\tilde r^4}\tilde\xi(\tilde r)+O(\epsilon_*^2)\ .\nonumber
\end{eqnarray}
The first two terms in (\ref{Gibbs1}) can be cancelled by appropriate counter terms (we refer the reader to the next section for more details), while the last term 
is precisely the DBI-term describing a probe D7--brane. Therefore to first order in $\epsilon_*\sim N_f/N_c$ the Gibbs free energy of the unquenched system coincides with the free energy calculated in the quenched approximation. This feature has been observed also in the finite temperature case studied in \cite{arXiv:0909.2865}. This suggests that to first order in $\epsilon_*$ one cannot study the effect of the dynamically generated energy scale set by the Landau pole of the dual gauge theory. Therefore we need to compute the second order contribution to the Gibbs free energy\footnote{Note that from a holographic point of view even at first order in $\epsilon_*$ our ansatz provides a novel feature, namely the relation between the field $J(r)$ and the magnetization of the dual gauge theory.}. Furthermore, despite the fact that we are dealing with a finite cut off set by the parameter $r_*$ we can still perform a holographic renormalization of the free energy resulting in an expression finite in the large $r_*$ limit.

%%%%%%%%%%%%%%%%%%%%%%%%%%%%%%%%%%%%%%%%%%%%%%%%%%%%%%%%%%%%%%%%%%%%%%%%%%%%%%%

\subsection{Holographic renormalization}

In this subsection we will regularize the Gibbs free energy defined in (\ref{Gibbsfree}). One can show, that modulo logarithmic divergences, the counter terms needed to regulate (\ref{Gibbsfree}) are the same as ones needed to regulate the supersymmetric background corresponding to vanishing magnetic field (the $r_m\to 0$ limit). Therefore, it is natural to regulate the on-shell action in (\ref{Gibbs1}) by subtracting the supersymmetric on-shell action. Note that this would suggest regulating the DBI contribution to the free energy by subtracting the DBI term for the supersymmetric case. Such an approach would make comparison to the results obtained in the quenched approximation difficult, where the on-shell action is regulated by the addition of the appropriate boundary terms. Furthermore, logarithmic divergences depending on the magnetic field do not have analogues in the supersymmetric background. To rectify this we consider a mixed regularization scheme: we regulate logarithmic and divergences due to the DBI term by adding appropriate covariant boundary counter terms, while the rest we directly cancel by subtracting the relevant part of the supersymmetric action.

%%%%%%%%%%%%%%%%%%%%%%%%%%%%%%%%%%%%%%%%%%%%%%%%%%%%%%%%%%%%

\subsubsection{The supersymmetric case}

In the supersymmetric limit of vanishing magnetic field, equations  (\ref{EOM-chi}) and (\ref{diff-b})-(\ref{diff-H}) can be integrated to give a BPS system of first order differential equations\footnote{Note that in this limit $b\equiv 1$ and $H\equiv J\equiv 0. $}. The full non-perturbative solution for the corresponding supersymmetric background was obtained in \cite{arXiv:0807.0298} and for more details we refer the reader to the appendix \ref{app-susy}. Here we present the expansion to second order in $\epsilon_*$ of the supersymmetric Euclidean ``on-shell action" and analyze its divergences. We obtain:
\begin{eqnarray}
-\frac{2\kappa_{10}^2}{V_{4}\pi^3}{\cal I}_{\rm susy}&=&6r_*^4+4\epsilon_*\left[r_*^4\eta(r_*)-\int\limits_{ m_0}^{r_*}dr\,r^3\, \cos^3\frac{\chi_0}{2}\sqrt{1+\frac{r^2}{4}\chi_0'^2}\right]+\label{SUSYaction}\\
&+&
\epsilon_*^2\left[ 4r_*^4\eta_2(r_*)-\frac{r^5\cos^3\frac{\chi_0}{2}\chi_0'}{\sqrt{1+\frac{r^2}{4}\chi_0'^2}}\chi_1\Big|_{r=r_*}-\frac{4}{5}r_*^4\left(4{\Delta_1^s(r_*)}^2+\Lambda_1^s(r_*)^2\right)\right .+\nonumber\\
&+&\left.\frac{4}{5}r_*^5\left(\Lambda_1^s\,{\Lambda_1^s}'-\Delta_1^s\,{\Delta_1^s}'\right)\Big|_{r=r_*}-\frac{1}{2}\int\limits_{m_0}^{r_*}drr^3\,\cos^8\frac{\chi_0}{2}-\frac{1}{2}\int\limits_{m_0}^{r_*}drr^5J_{\Phi}^s\Phi^s+\right. \nonumber\\
&+&\left. \frac{4}{5}\int\limits_{m_0}^{r_*}drr^5\left(J_{\Lambda_1}^s\Lambda_1^s-J_{\Delta_1^s}\Delta_1^s\right) \right]+O(\epsilon_*^3)\ .\nonumber
\end{eqnarray}
As we pointed out above the divergences of the action (\ref{SUSYaction}) completely cancel (modulo logarithmic terms) those of the non-supersymmetric action. However we would like to regulate together the DBI term in the first order expansion and the term depending on $\chi_1$ in the second order expansion (coming from expanding $\chi=\chi_0+\epsilon_*\chi_1$ in the DBI term) by adding covariant counter-terms. We make this choice  in order for our regularization scheme to be compatible with the one employed in \cite{Albash:2007bk}, for the probe approximation.  Furthermore, by construction, boundary terms which do not contain derivatives of the fields are exactly the same with those of the non-supersymmetric action and can be directly subtracted. We define a subtracting action:
\begin{equation} 
{\cal I}_{\rm subt}\equiv {\cal I}_{\rm subt}^{\rm bdr}+{\cal I}_{\rm subt}^{\rm bulk}\ ,\label{Isubt}
\end{equation}
where:
\begin{eqnarray}
\frac{2\kappa_{10}^2}{V_{4}\pi^3}{\cal I}_{\rm subt}^{\rm bdr}&\equiv&6r_*^4+4\epsilon_* r_*^4\eta(r_*)+\epsilon_*^2\left[ 4r_*^4\eta_2(r_*)-\frac{4}{5}r_*^4\left(4{\Delta_1^s(r_*)}^2+\Lambda_1^s(r_*)^2\right)\right ]\ ,\\
\frac{2\kappa_{10}^2}{V_{4}\pi^3}{\cal I}_{\rm subt}^{\rm bulk}&\equiv&\epsilon_*^2\left[ \frac{4}{5}r_*^5\left(\Lambda_1^s\,{\Lambda_1^s}'-\Delta_1^s\,{\Delta_1^s}'\right)\Big|_{r=r_*}-\frac{1}{2}\int\limits_{m_0}^{r_*}drr^3\,\cos^8\frac{\chi_0}{2}-\frac{1}{2}\int\limits_{m_0}^{r_*}drr^5J_{\Phi}^s\Phi^s+\right. \nonumber\\
&+&\left. \frac{4}{5}\int\limits_{m_0}^{r_*}drr^5\left(J_{\Lambda_1}^s\Lambda_1^s-J_{\Delta_1^s}\Delta_1^s\right) \right]=-\epsilon_*^2\frac{(r_*^2-m^2)^5(11r_*^2+7m^2)}{162r_*^8}\ ,
\end{eqnarray}
and we have explicitly evaluated ${\cal I}_{\rm subt}^{\rm bulk}$. Next we proceed with the regularization of:
\begin{eqnarray}
\frac{2\kappa_{10}^2}{V_{4}\pi^3}({\cal I}_{\rm susy}+{\cal I}_{\rm subt})&=&4\epsilon_*\int\limits_{ m_0}^{r_*}dr\,r^3\, \cos^3\frac{\chi_0}{2}\sqrt{1+\frac{r^2}{4}\chi_0'^2}+\epsilon_*^2\frac{r^5\cos^3\frac{\chi_0}{2}\chi_0'}{\sqrt{1+\frac{r^2}{4}\chi_0'^2}}\chi_1\Big|_{r=r_*}+{\cal O} (\epsilon_*^3)
\nonumber \\
&=&4\epsilon_*\int\limits_{ m_0}^{r_*}dr\,r^3\, \cos^3\frac{\chi}{2}\sqrt{1+\frac{r^2}{4}\chi'^2} + {\cal O} (\epsilon_*^3)  \,  .
\label{susysubtr}
\end{eqnarray}
Note that in terms of the full function $\chi(r)$ the subtracted supersymmetric action looks just like the DBI term of the probe approximation. However, the function $\chi(r)$ is not the same (namely $2\arcsin(m/r) $). 
\paragraph{}
The $\kappa$-symmetry condition for the embedding can be solved in a new radial coordinate $\hat r(r)$:
\begin{equation}
\hat r(r)=r\left(1+\epsilon_*\rho_1(r)+O(\epsilon_*^2)\right)\ .
\end{equation}
where $\rho_1(r)$ is given in (\ref{rho1}),  with $\rho_1(r_*)=0$ and hence $\hat r(r_*)=r_*$. It is in this new radial coordinate that the fiducial embedding satisfies:
\begin{equation}
\chi(\hat r)=2\arcsin\frac{\hat m}{\hat r}\ ,
\end{equation}
where $\hat m=\hat r(m_0)$. In fact we identify $\hat m$ as the full bare mass parameter $\hat m\equiv\hat r(m_0)=m_0+\epsilon_*m_1+O(\epsilon_*^2)$. Clearly this defines $m_1$ in terms of $m_0$. Furthermore, since $\hat r(r_*)=r_*$ one can write:
\begin{equation}
\chi(r_*)=\frac{2m}{r_*}+ {\cal O}\left(\frac{1}{r_*^3}\right)=\chi_0(r_*)+\epsilon_*\chi_1(r_*)+
 {\cal O}\left(\epsilon_*^2\right)=\frac{2m_0}{r_*}+\epsilon_*\chi_{1}(r_*)+ {\cal O}\left(\frac{1}{r_*^3}\right)+ {\cal O}(\epsilon_*^2)\ ,
\end{equation}
suggesting the natural relation between $m_1$ and the leading term in the expansion of $\chi_1(r_*)$:
\begin{equation}
\chi_1(r_*)=\frac{2m_1}{r_*}+ {\cal O}\left(\frac{1}{r_*^3}\right)\ . \label{bare mass1}
\end{equation}
Equation (\ref{bare mass1}) is an alternative definition of the first correction $m_1$ to the bare mass, which does not require the existence of a special radial variable $\hat r$. We will use this definition when we study the theory at finite magnetic field. 

The fact that at $r_*$ the fiducial embedding  still has the same expansion as in the quenched approximation suggests that one can use the same counter terms. 
Indeed one can check that the following counter terms:
\begin{equation}
\frac{-2\kappa_{10}^2}{R^4V_{4}\pi^3}\frac{{\cal I}_{\rm count}}{\sqrt{\gamma}}=\epsilon_*\left(1-\frac{\chi^2}{2}+\frac{5\chi^4}{48}\right)_{r_*}=\epsilon_*\left(1-\frac{\chi_0^2}{2}+\frac{5\chi_0^4}{48}\right)_{r_*}+\epsilon_*^2\left(-{\chi_0}\chi_1+\frac{5\chi_0^3\chi_1}{12}\right)_{r_*}+O(\epsilon_*^3)\ ,\label{Icount}
\end{equation}
completely cancel the first order term in (\ref{susysubtr}) and leaves only sub-leading terms at second order in $\epsilon_*$. More precisely:
\begin{equation} 
{\cal I}_{\rm susy}+{\cal I}_{\rm subt}+{\cal I}_{\rm count}=0+\epsilon_*^2\left[0+
{\cal O} \left(\#(m)\frac{\log r_*}{r_*^2}\right)\right]+{\cal O}\left(\epsilon_*^3\right)\ .\label{susyregreg}
\end{equation}
If the theory were UV complete the addition of the regulating terms ${\cal I}_{\rm subt}$ and ${\cal I}_{\rm count}$ to the ``on-shell" action would provide a vanishing mass dependent expression for the free energy of the supersymmetric background. However, due to the existence of a Landau pole our UV cutoff $r_*$ is finite. In general this can lead to spurious mass dependence of the free energy (and hence spurious condensate) due to the $\#(m)$ term in (\ref{susyregreg}). Fortunately a similar contribution from sub-leading terms is present in the non-supersymmetric (finite magnetic field) case too. One can verify that the two contributions are approximately equal for large bare masses, which is also the regime when these contributions are significant. Therefore, ignoring sub-leading terms is a very good approximation, because they can always be canceled by a redefinition of ${\cal I}_{\rm subt}$. In the next subsection we will regularize the free energy at finite magnetic field by adding ${\cal I}_{\rm subt}+{\cal I}_{\rm count}$ to the action $\tilde {\cal I}$.

%%%%%%%%%%%%%%%%%%%%%%%%%%%%%%%%%%%%%%%%%%%%%%%%%%%%%%%%%%%%%%%%%%%%%%%%%%%%%%%%%

\subsubsection{Regularization at finite magnetic field}

In this subsection we regularize the ``on-shell" action  $\tilde {\cal I}$ by adding the term ${\cal I}_{\rm subt}$ defined in (\ref{Isubt}) and the counter terms ${\cal I}_{\rm count}$ 
defined in (\ref{Icount}). It is convenient to split the contributions to $\tilde{\cal I}$ into two parts $\tilde{\cal I}=\tilde {\cal I}_{\chi}+\tilde {\cal I}_{\rm bulk}$, regulated by ${\cal I}_{\rm count}$ and ${\cal I}_{\rm subt}$ respectively. We find:
\begin{eqnarray}
&&\frac{2\kappa_{10}^2}{V_{4}\pi^3}(\tilde{\cal I}_{\rm bulk}+{\cal I}_{\rm subt})=\epsilon_*^2r_m^4\left[\int\limits_{\tilde r_q}^{\tilde r_*}d\tilde r\tilde r^5
\left[\frac{1}{4}J_{b_1}b_1+\frac{1}{2}J_{\lambda_1}\lambda_1-\frac{4}{5}\left(J_{\Lambda_1}\Lambda_1-J_{\Delta_1}\Delta_1\right)+\frac{1}{2}J_{\Phi}\Phi\right]
+\right.\label{actionbulk} \\
&&\left. +\frac{1}{2}\int\limits_{\tilde r_q}^{\tilde r_*}d\tilde r\tilde r\left[J_{H_1}H_1+\left(\tilde r^2+\frac{1}{\tilde r^2}\right)\cos^8\frac{\chi_0}{2}\right]- \frac{4}{5}\tilde r_*^5\left(\Lambda_1\,{\Lambda_1}'-\Delta_1\,{\Delta_1}'\right)\Big|_{\tilde r=\tilde r_*}\right]-\epsilon_*^2\frac{(r_*^2-m_0^2)^5(11r_*^2+7m_0^2)}{162r_*^8}\ .\nonumber
\end{eqnarray}
The expression in equation (\ref{actionbulk}) can be evaluated numerically and analytically in the limit $r_m\ll m_0$ (weak magnetic field). In the limit of weak magnetic field one finds that the subtracting term regulates all divergences. We will study this in more details in the next subsection. Let us now focus on the remaining part of the action:
\begin{eqnarray}
\frac{2\kappa_{10}^2}{V_{4}\pi^3}(\tilde{\cal I}_{\rm \chi}+{\cal I}_{\rm count})&=&\epsilon_*\left[4\,r_m^4\int\limits_{ \tilde r_q}^{\tilde r_*}d\tilde r\,\tilde r\,\sqrt{1+\tilde r^4}\, \cos^3\frac{\chi_0}{2}\sqrt{1+\frac{\tilde r^2}{4}\chi_0'^2}-\tilde r_*^4\left(1-\frac{\chi_0^2}{2}+\frac{5\chi_0^4}{48}\right)_{\tilde r_*}\right]+\nonumber\\
&+&\epsilon_*^2r_m^4\left[\frac{\tilde r^3\sqrt{1+\tilde r^4}\cos^3\frac{\chi_0}{2}\chi_0'}{\sqrt{1+\frac{\tilde r^2}{4}\chi_0'^2}}\Big|_{\tilde r=\tilde r_*}+\tilde r_*^4\left(-{\chi_0}+\frac{5\chi_0^3}{12}\right)_{r_*}\right]\chi_1(\tilde r_*)\ .
\end{eqnarray}
The first term in the first order contribution to the action ${\cal I}_{\rm \chi}+{\cal I}_{\rm count}$ is the same as in the quenched approximation and hence one needs to complete the counter term action to subtract the $\propto r_m^4\log r_*$ divergence. The complete counter term action is:
\begin{equation}
\frac{2\kappa_{10}^2}{V_{4}\pi^3}\tilde {\cal I}_{\rm count}=\frac{2\kappa_{10}^2}{V_{4}\pi^3}{\cal I}_{\rm count}-\epsilon_*\,B_{\alpha\beta}B^{\alpha\beta}\Big |_{r_*}\log\frac{r_*}{R}\ .
\end{equation}
Now using the fact that $\chi_0(\tilde r_*)=2\tilde m_0/\tilde r_*+2\tilde c_0/\tilde r_*^3+O(1/\tilde r_*^5)$ one obtains:
\begin{equation} 
\frac{2\kappa_{10}^2}{V_{4}\pi^3}(\tilde{\cal I}_{\rm \chi}+\tilde{\cal I}_{\rm count})=4\epsilon_*\,r_m^4\,\left[\tilde {\cal I}_{\rm D7}(\tilde m_0)+\epsilon_*\left(-2\tilde c_0\,\tilde m_1+O\left(\frac{\log \tilde r_*}{\tilde r_*^2}\right)\right)\right]+O(\epsilon_*^3)\ ,\label{Ichireg}
\end{equation}
where we have defined $\tilde {\cal I}_{\rm D7}(\tilde m)$ and $\tilde m_1$ via the following relations:
\begin{eqnarray}
\tilde{\cal I}_{\rm D7}(\tilde m_0)&=&\int\limits_{\tilde r_q}^{\tilde r_*}d\,\tilde r\, \tilde r \sqrt{1+\tilde r^4}\tilde\xi(\tilde r)-\frac{\tilde r_*^4}{4}\left(1-\frac{\chi_0^2}{2}+\frac{5\chi_0^4}{48}\right)_{\tilde r_*}-\frac{1}{2}\log\frac{r_*}{R}\ ,\label{ID7} \\
\chi_1(\tilde r_*)&=&\frac{2\tilde m_1}{\tilde r_*}+O\left(\frac{\log\tilde r_*}{\tilde r_*^3}\right)\ .
\end{eqnarray}
The function $\tilde {\cal I}_{\rm D7}(\tilde m_0)$ is proportional to the expression for the free energy of the fundamental matter in the quenched approximation.  It is known that (see for example \cite{Mateos:2007vn}) ${d\tilde{\cal I}_{\rm D7}(\tilde m_0)}/{d\tilde m_0}=-2\tilde c_0$. Using this relation (\ref{Ichireg}) can be written as:
\begin{eqnarray}
\frac{2\kappa_{10}^2}{V_{4}\pi^3}(\tilde{\cal I}_{\rm \chi}+\tilde{\cal I}_{\rm count})&=&4\epsilon_*\,r_m^4\,\left[\tilde {\cal I}_{\rm D7}(\tilde m_0)+\tilde{\cal I}_{\rm D7}'(\tilde m_0)(\epsilon_*\, \tilde m_1)+\epsilon_*\left(O\left(\frac{\log \tilde r_*}{\tilde r_*^2}\right)\right)\right]+O(\epsilon_*^3)= \nonumber\\
&=&4\epsilon_*\,r_m^4\,\tilde {\cal I}_{\rm D7}(\tilde m)+\epsilon_*^2\left(O\left(\frac{\log \tilde r_*}{\tilde r_*^2}\right)\right)+O(\epsilon_*^3)\ ,
\end{eqnarray}
where we have defined the complete bare mass parameter as $m=m_0+\epsilon_*\,m_1+O(\epsilon_*^2)$. Note that modulo sub-leading terms the only contribution from the $\epsilon_*^2$ term in (\ref{Ichireg}) is the one needed to complete the argument of $\tilde {\cal I}_{\rm D7}$. Now we can write our final expression for the regularized action $\tilde {\cal I}_{\rm reg}\equiv\tilde{\cal I}+\tilde{\cal I}_{\rm subt}+\tilde{\cal I}_{\rm count}$:
\begin{equation}
\frac{2\kappa_{10}^2}{V_{4}\pi^3}\tilde{\cal I}_{\rm reg}=4\epsilon_*\,r_m^4\,\tilde {\cal I}_{\rm D7}(\tilde m)+\epsilon_*^2\,r_m^4\tilde {\cal I}_{\rm D7}^{(2)}(\tilde m)+O(\tilde\epsilon_*^3)\ ,\label{regAction}
\end{equation}
where the quantity $\tilde {\cal I}_{\rm D7}^{(2)}(\tilde m)$ is given by the right-hand side of  (\ref{actionbulk}) divided by $\epsilon_*^2\, r_m^4$ and we have completed its argument to a full bare mass $m$. We are now ready to proceed with the evaluation of the free energy and the fundamental condensate as a function of the bare mass. We will study first the limit of weak magnetic field $(\tilde m\gg 1)$, which can be analyzed analytically.

%%%%%%%%%%%%%%%%%%%%%%%%%%%%%%%%%%%%%%%%%%%%%%%%%%%%%%%%%%%%%%%%%%%%%%%%%%%%%%%%%

\subsection{Free energy and condensate at weak magnetic field}

In this section we calculate analytically the Gibbs free energy ${\cal G}$ and the fundamental condensate of the theory in the limit of weak magnetic field. Given that $H_*\propto r_m^2$ and $\tilde m=m/r_m$ this corresponds to the $1/\tilde m$ expansion in equation (\ref{regAction}). It has been shown in \cite{Albash:2007bk} that:
\begin{equation}
\tilde{\cal I}_{\rm D7}(\tilde m)=-\frac{1}{2}\log\tilde m+{\rm const_1}+O(1/\tilde m^4)\ . 
\end{equation}
Furthermore, using the approximate analytic solution for the background (see Appendix \ref{WMF}) one can show that:
\begin{equation}  
\tilde{\cal I}_{\rm D7}^{(2)}(\tilde m)={\rm const_2 }+O\left(\frac{\log (\tilde r_*/\tilde m)}{\tilde m^4}\right)\ .\label{Actsecweak}
\end{equation}
Note that, since we are dealing with a finite cutoff, numerically, the logarithmic term in (\ref{Actsecweak}) is of order one and hence 
the sub-leading term is of order $1/\tilde m^4$. All together we obtain:
\begin{equation}
\frac{2\kappa_{10}^2}{V_{4}\pi^3}\tilde{\cal I}_{\rm reg}(\tilde m)=-2\epsilon_*(\,r_m^4\log\tilde m+{\rm const})+O(\epsilon_*^3)\propto -\epsilon_*[2H_*^2\log m +\#( H_*)]+O(\epsilon_*^3)\ ,
\end{equation}
where we have ignored terms of order $(1/\tilde m^4)$. For the leading order contribution to the free energy we recover the result from the quenched approximation:
\begin{equation}
\frac{1}{V_3}{{\cal G}(H_*,m)}=-\epsilon_*\frac{\pi^3}{\kappa_{10}^2}r_m^4\log\,m+O(\epsilon_*^3)\ ,
\end{equation}
where we have skipped the mass independent term $\#(H_*)$. Now we can define the fundamental condensate via:
\begin{equation}
{\langle\bar\psi\psi\rangle}\equiv\frac{1}{V_3}\frac{\partial{\cal G}(H_*,m)}{\partial m}=-\epsilon_*\,\frac{\pi^3}{\kappa_{10}^2}\frac{r_m^4}{m}+O(H_*^4)+O(\epsilon_*^3)\ .\label{condWeak}
\end{equation}
Equation (\ref{condWeak}) suggests that to order $\epsilon_*^3$ the fundamental condensate is the same as in the quenched approximation. However, as commented in section 4.6, the relevant perturbative parameter which takes into account the running of the effective coupling is $\epsilon_q$. Using (\ref{epsstar}) we arrive at the following result for the condensate:
\begin{equation}
\frac{\kappa_{10}^2}{\pi^3}{\langle\bar\psi\psi\rangle}=-\epsilon_q\,\frac{r_m^4}{m}\left(1+\epsilon_q\left(\frac{3}{4}+\log\frac{r_*}{m} \right)\right)+O(\epsilon_q^3)\ ,
\end{equation}
where we have used that for $\tilde r_q\gg1$ to leading order one has:
\begin{equation}
 \Phi_1(\tilde r_q)\approx-\frac{3}{4}-\log\frac{r_*}{r_q}\approx-\frac{3}{4}-\log\frac{r_*}{m}\ .
\end{equation}
Therefore our conclusion is that at first order in $\epsilon_q$ the result is the same as in the quenched approximation. However, at second order in $\epsilon_q$ we observe a logarithmic running of the condensate as a function of the bare mass and the energy scale set by the finite UV cutoff $\Lambda_{\rm UV}\sim r_*$, which reflects the positive beta function of the theory. 
We proceed with calculating the Gibbs free energy and the fundamental condensate at strong magnetic field.

%%%%%%%%%%%%%%%%%%%%%%%%%%%%%%%%%%%%%%%%%%%%%%%%%%%%%%%%%%%%%%%%%%%%%%%%%%%%%%%%%

\subsection{Free energy and condensate at strong magnetic field}

\begin{figure}[h] %  figure placement: here, top, bottom, or page
   \centering
   \includegraphics[width=3.4in]{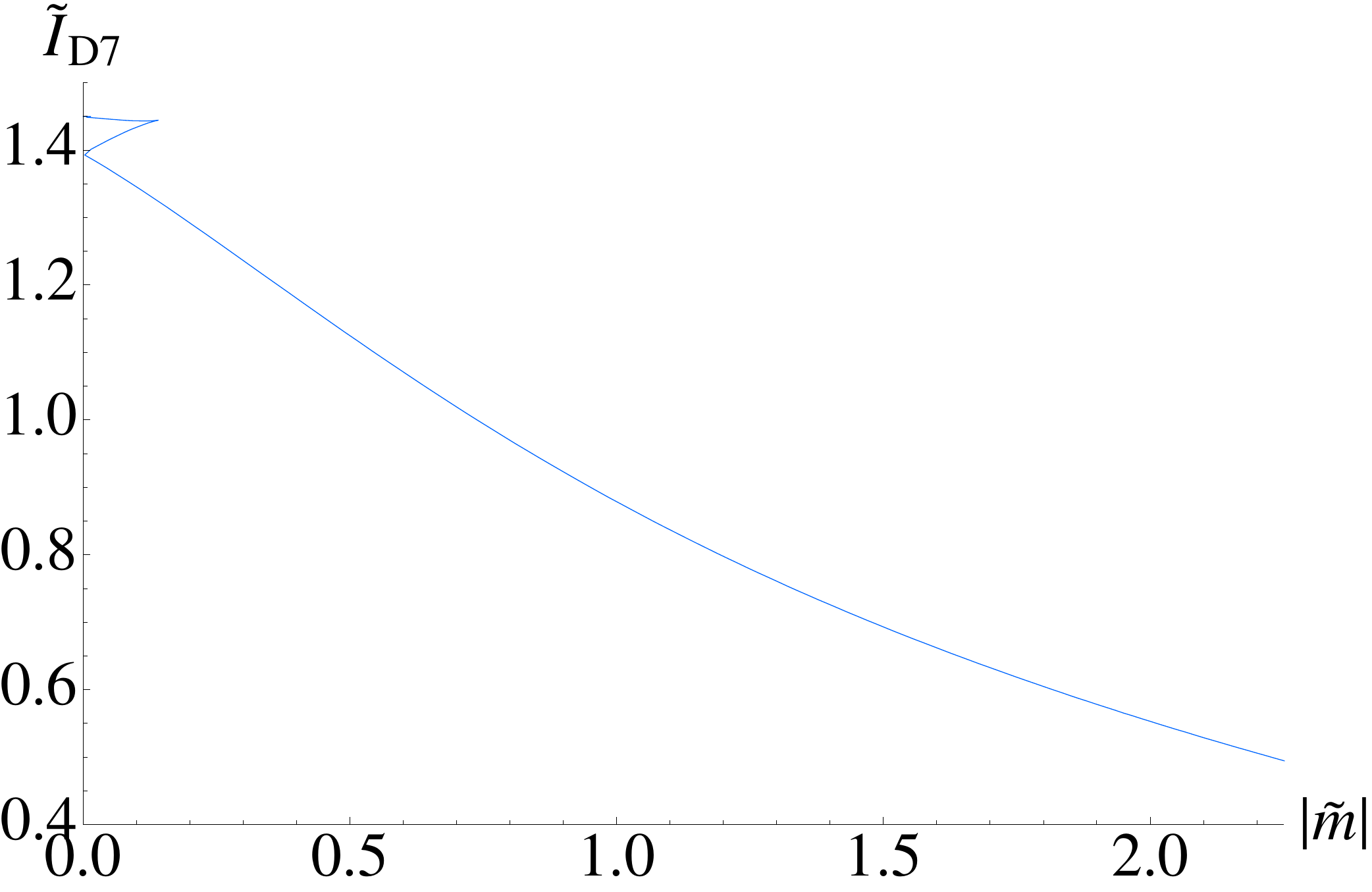} 
   \includegraphics[width=3.4in]{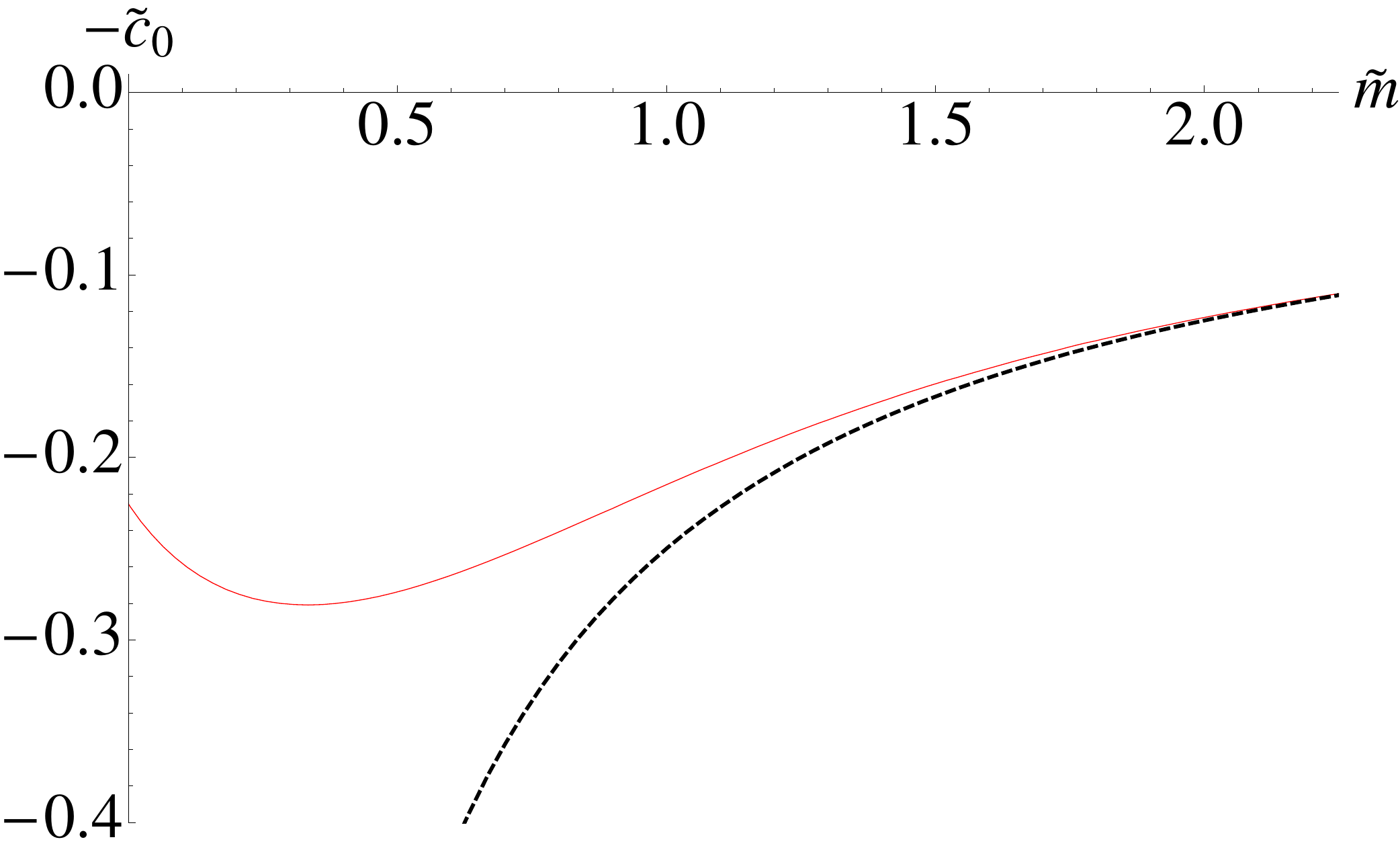} 
  \caption{Plots of  $\tilde{\cal I}_{\rm D7}$ and $\tilde c_0$ versus the bare mass parameter $\tilde m$.}
   \label{fig:fig1}
\end{figure}

In this subsection we calculate the Gibbs free energy and the condensate of the theory at strong magnetic field evaluating numerically the functions $\tilde{\cal I}_{\rm D7}(\tilde m)$ and $\tilde{\cal I}_{\rm D7}^{(2)}(\tilde m)$. As commented above the first function describes the Gibbs free energy to first order in $\epsilon_*$ and has been studied in  \cite{Erdmenger:2007bn,Albash:2007bk}. Apparently it also describes the first order contribution to the fundamental condensate given by:
\begin{equation}
\frac{\kappa_{10}^2}{2\pi^3r_m^3}\langle\bar\psi\psi\rangle= \epsilon_*\tilde{\cal I}_{\rm D7}'(\tilde m)+O(\epsilon_*^2)=-2\epsilon_*\,\tilde c_0(\tilde m)+O(\epsilon_*^2)\ .
\end{equation}
In figure \ref{fig:fig1} we present plots of $\tilde{\cal I}_{\rm D7}(\tilde m)$ and $-\tilde c_0(\tilde m)$.
As one can see from the plot of $\tilde{\cal I}_{\rm D7}(\tilde m)$ for sufficiently small $\tilde m$ the theory has multiple phases. However only the lowest branch is thermodynamically stable and studies of the meson spectrum in the quenched approximation \cite{Filev:2007qu} verified this stability. In the second plot of Figure {\ref{fig:fig1}} we see the condensate, but 
only for the stable phase. At vanishing bare mass the theory has a negative condensate, while for large bare masses the condensate has the $-1/(4\tilde m)\propto H_*^2/m$ dependence governing the weak magnetic field limit. 

\paragraph{}

Our next task is to obtain similar plots for the second order contribution in $\epsilon_*$ to the Gibbs free energy and the fundamental condensate. We will restrict ourselves 
in studying only the stable (to first order in $\epsilon_*$) phase of the theory. 
Using the numerical solution for the background we study $\tilde{\cal I}_{\rm D7}^{(2)}(\tilde m)$ and its derivative. Note that (\ref{regAction}) and the definition of the condensate:
\begin{equation}
\langle\bar\psi\psi\rangle\equiv\frac{1}{V_3}\frac{\partial{\cal G}(H_*,m)}{\partial m}=\frac{1}{V_4}\frac{\partial{\tilde{\cal I}}_{\rm reg}(H_*,m)}{\partial m}\ ,
\end{equation}
imply the relation:
\begin{equation}
\frac{\kappa_{10}^2}{2\pi^3r_m^3}\langle\bar\psi\psi\rangle= \epsilon_*\tilde{\cal I}_{\rm D7}'(\tilde m)+\frac{1}{4}\epsilon_*^2{\tilde{\cal I}_{\rm D7}^{(2)}}{}'(\tilde m)+O(\epsilon_*^3)=-2\epsilon_*\,\tilde c_0(\tilde m)-2\epsilon_*^2\tilde c_2(\tilde m)+O(\epsilon_*^3)\ ,\label{condfull}
\end{equation}
where $\tilde c_2$ is defined by ${\tilde{\cal I}_{\rm D7}^{(2)}}{}'(\tilde m)=-8\tilde c_2(\tilde m) $. 
\begin{figure}[h] %  figure placement: here, top, bottom, or page
   \centering
   \includegraphics[width=3.4in]{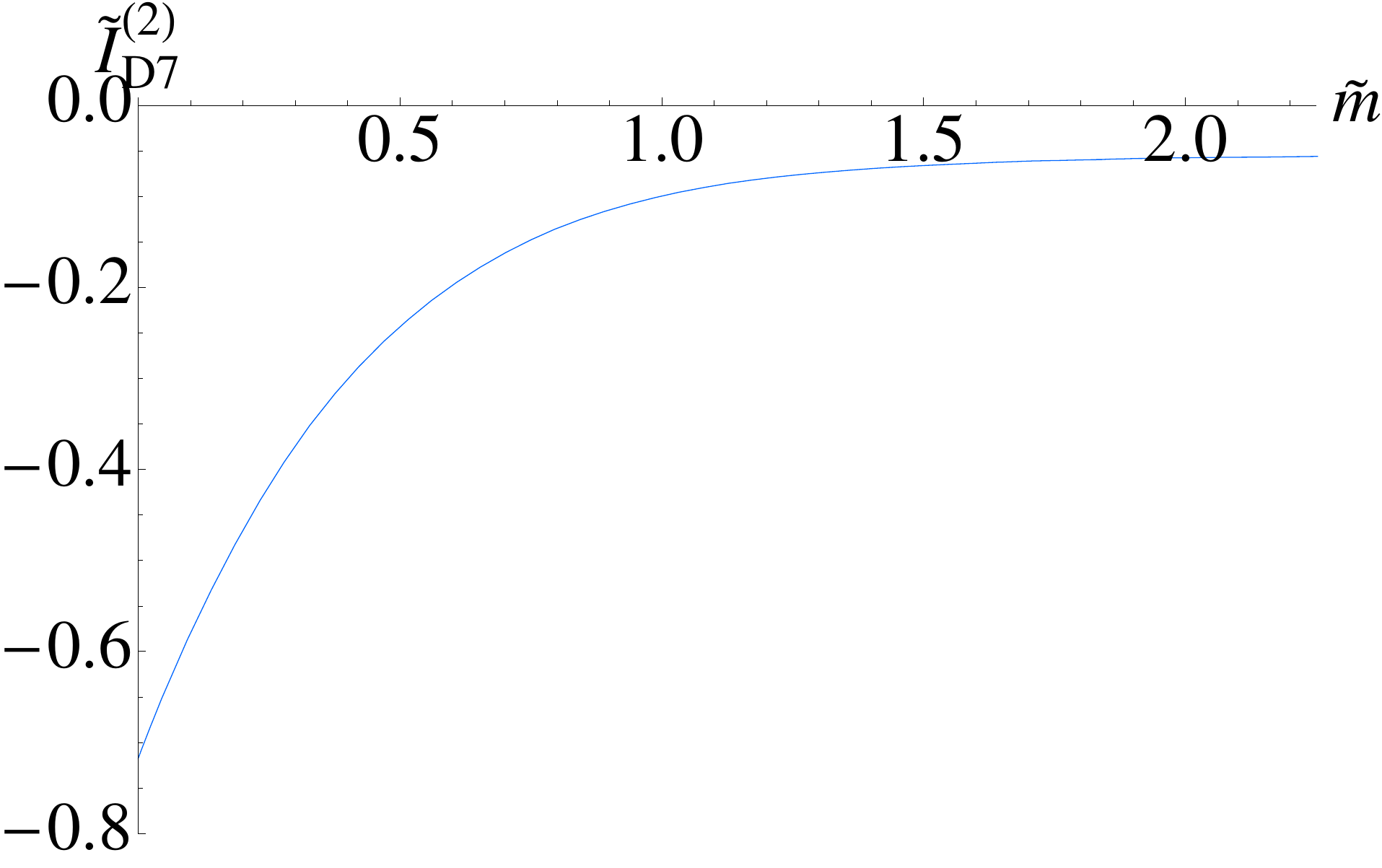} 
   \includegraphics[width=3.4in]{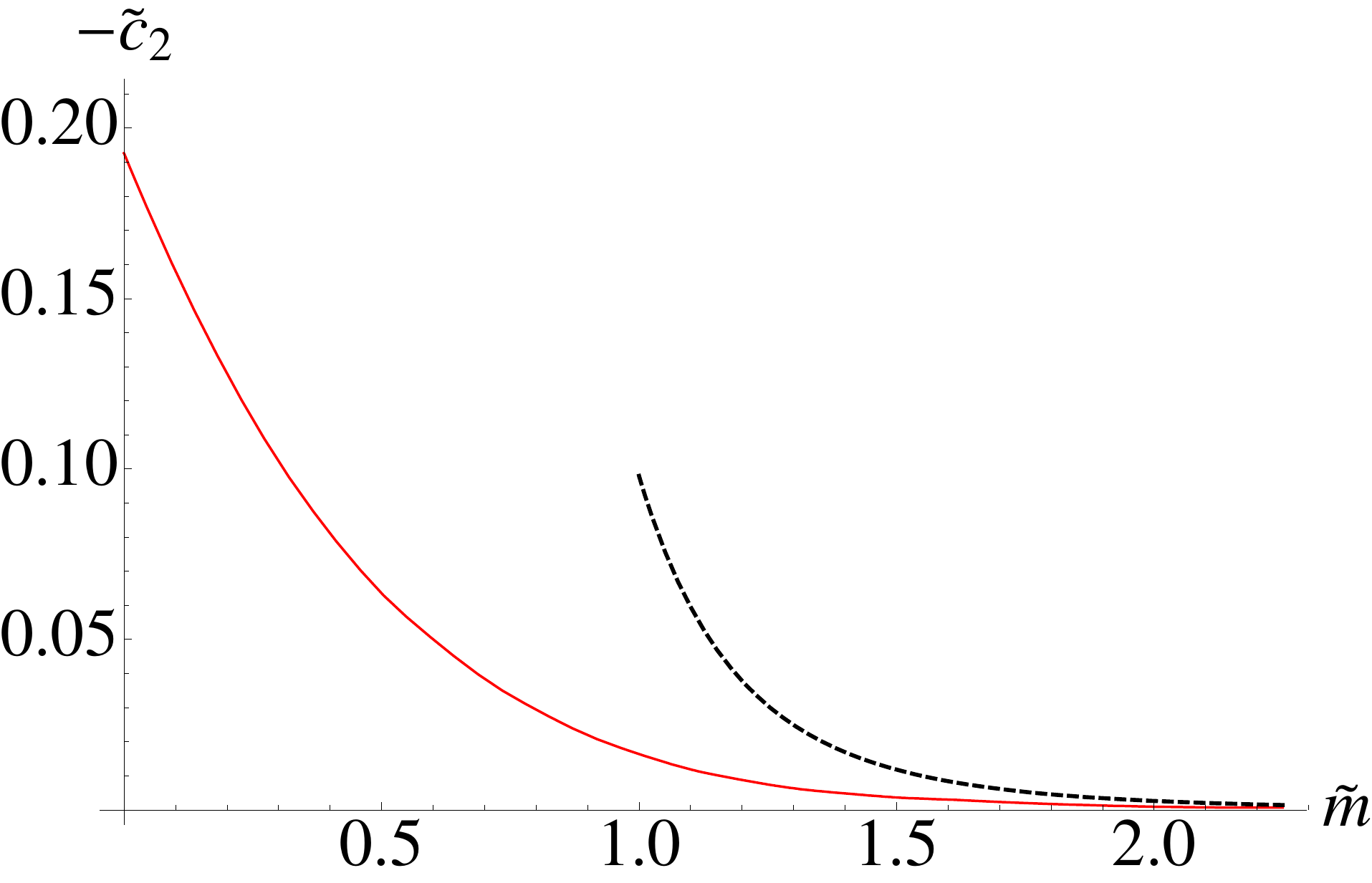} 
  \caption{Plots of  $\tilde{\cal I}_{\rm D7}^{(2)}$ and $-\tilde c_2$ versus the bare mass parameter $\tilde m$.}
   \label{fig:fig2}
\end{figure}
The corresponding plots are presented in figure \ref{fig:fig2} and from these one can see that $\tilde c_2$ approaches zero much faster than $\tilde c_0$ in Figure \ref{fig:fig1}. Surprisingly the second order contribution to the fundamental condensate is positive. However the relevant perturbative parameter is $\epsilon_q$ (defined in section 2.6) therefore in order to obtain the second order correction to the condensate we need to trade $\epsilon_*$ for $\epsilon_q$ in (\ref{condfull}). Using (\ref{epsstar}) and the fact that $\Phi_1(\tilde r_q)<0$ (see Figure \ref{fig:fig0}) we obtain:
\begin{eqnarray}
\frac{\kappa_{10}^2}{2\pi^3r_m^3}\langle\bar\psi\psi\rangle&=& -2\epsilon_q\,\tilde c_0(\tilde m)-2\epsilon_q^2(\tilde c_2(\tilde m)+|\Phi(\tilde r_q)|\tilde c_0(\tilde m ))+O(\epsilon_q^3)\label{condfullQ}\\
&=& -2\epsilon_q\,\tilde c_0(\tilde m)-2\epsilon_q^2(\hat c_2(\tilde m))+O(\epsilon_q^3)\ , \nonumber
\end{eqnarray}
where we have defined:
\begin{equation}
\hat c_2(\tilde m)\equiv\tilde c_2(\tilde m)+|\Phi(\tilde r_q)|\tilde c_0(\tilde m )\ .
\end{equation}

As one can see from figure \ref{fig:fig0}$, |\Phi(\tilde r_q)|>1$. Furthermore, comparing Figure \ref{fig:fig1} and Figure \ref{fig:fig2} one can see that $|\tilde c_0(\tilde m)|>|\tilde c_2(\tilde m)|$. Therefore we conclude that $-\hat c_2(\tilde m)<0$ and the second order correction to the condensate is negative. This implies that taking into account internal fundamental loops in the calculation of the condensate does not change qualitatively the results obtained in the quenched approximation. 
\begin{figure}[h] %  figure placement: here, top, bottom, or page
   \centering
   \includegraphics[width=7in]{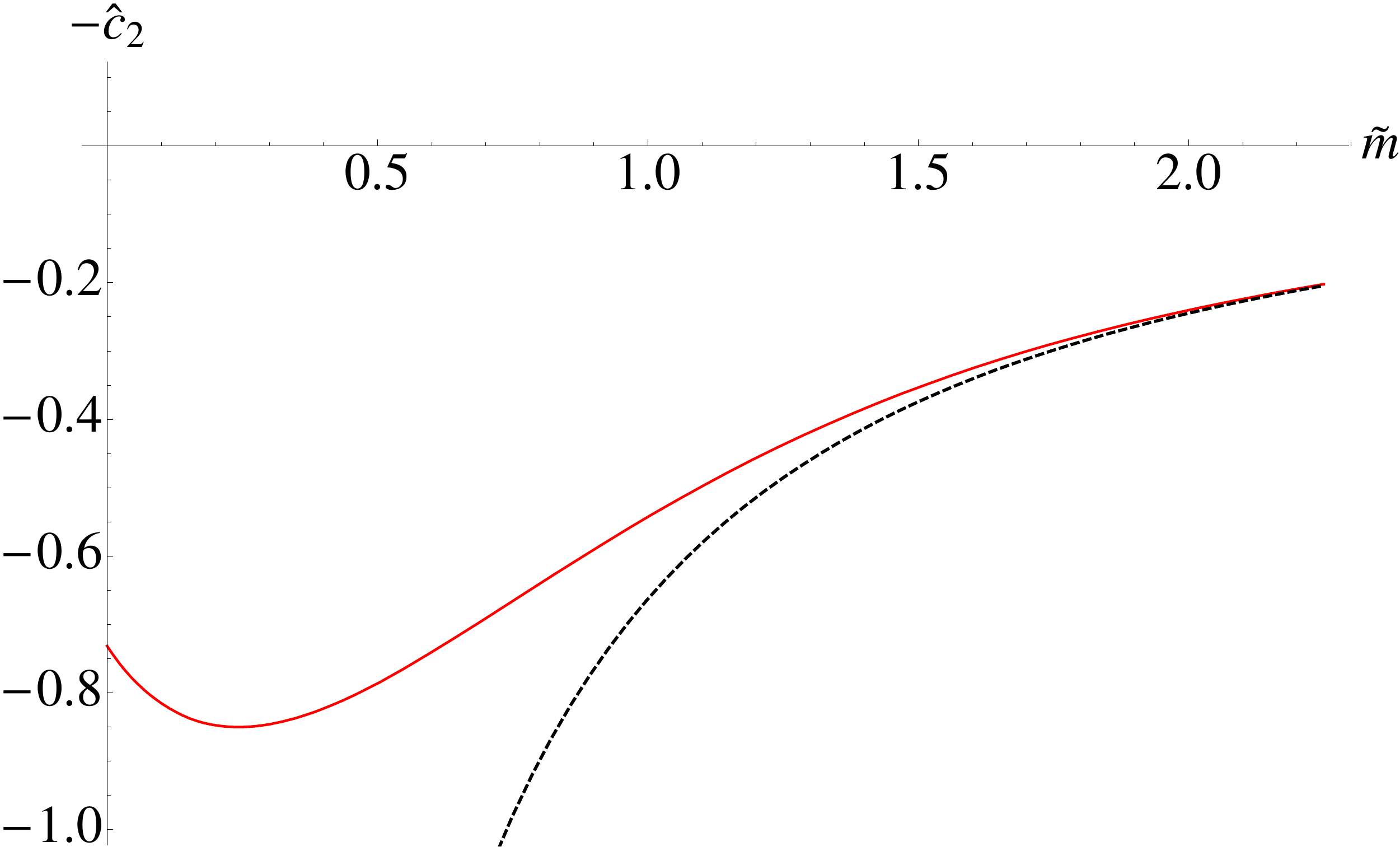} 
  \caption{A plot of $-\hat c_2$ versus $\tilde m$.}
   \label{fig:fig3}
\end{figure}
In fact it is not difficult to obtain a plot of $\hat c_2(\tilde m)$. Our results are presented in Figure \ref{fig:fig3}. One can see that indeed the second order contribution to the fundamental condensate is negative. The black dashed curve represents the function $\frac{1}{4\tilde m}(\frac{3}{4}+\log\frac{\tilde r_*}{\tilde m})$, which governs the asymptotic behavior of the condensate at large $\tilde m$ (weak magnetic field). 

Overall our conclusion is that  to leading order in a perturbative expansion in the ratio $N_f/N_c$, the free energy and fundamental condensate of the theory agree with the results obtained in the quenched approximation. Furthermore, at second order in  $N_f/N_c$ we observe that the effect of magnetic catalysis is enhanced and the contribution to the condensate of the theory from internal fundamental loops runs logarithmically with the finite cutoff $\Lambda_{UV}$.

%%%%%%%%%%%%%%%%%%%%%%%%%%%%%%%%%%%%%%%%%%%%%%%%%%%%%%
\section{Summary \& Conclusions}
Let us briefly summarize our conclusions and outline possible directions for future studies:
\paragraph{} 
In section 2 of this paper we have constructed a backreacted supergravity background holographically dual to an SU($N_c$) ${\cal N}=4$ SYM theory coupled to $N_f$ flavours of ${\cal N}=2$ hypermultiplets in the presence of external magnetic field $H_*$.  Our solution is perturbative in a parameter which counts 
the number of internal fundamental loops. At small distances the geometry has a hollow cavity, where it is similar to the supergravity dual of non-commutative SYM theory. The radius of this cavity, $r_q$, sets the energy scale corresponding to the physical mass of the fundamental fields. From holographic point of view,  the supergravity solution inside the cavity corresponds to the low energy effective field theory obtained after integrating out the massive flavour fields. The non-commutative nature of the theory reflects that fact that the fundamental fields are coupled to an external magnetic field, $H_*$. We have shown that the parameter of non-commutative along the plane perpendicular to the magnetic field, $\Theta^{23}$, scales as $\Theta^{23}\sim \frac{N_f}{N_c}\frac{1}{H_*}$.  
\paragraph{} 
At large radial distances,  our solution is well-approximated by the
non-perturbative supersymmetric solution corresponding to a vanishing
magnetic field \cite{arXiv:0807.0298}.  We introduce a sufficiently
large radial parameter ($r_*\gg r_q$) at which we match our
perturbative solution to the supersymmetric one. In the
holographically dual field theory, the radial parameter $r_*$
corresponds to a finite cutoff $\Lambda_{UV}$. At a given radial
distance $r_{LP}$  ($r_{LP}\gg r_*$ ), the dilaton field diverges, reflecting the positive $\beta$-function of the dual gauge theory and the existence of a Landau pole at an energy scale, $\Lambda_{LP}$ ($\Lambda_{LP}\sim r_{LP}$). We have the following relation \cite{arXiv:0807.0298}, $\epsilon_*\propto \log\frac{\Lambda_{LP}}{\Lambda_{UV}}$ implying that to keep the finite cutoff sufficiently far away from the Landau pole we need to keep the parameter $\epsilon_*\sim \frac{N_f}{N_c}$ sufficiently low.  Clearly in the limit $\epsilon_*\to 0$ we can make the cutoff $\Lambda_{UV}$ arbitrarily large, on field theory side this corresponds to the quenched approximation. In fact our studies of the free energy and the fundamental condensate of the dual gauge theory confirm that at leading order in $\epsilon_*$, we reproduce the results of the quenched approximation. 
\paragraph{} 
In section 3 we have applied our construction to the phenomenon of
magnetic catalysis, i.e. of mass generation in an external magnetic
field, beyond the quenched approximation. Note that our studies are at
fixed value of the external magnetic field, $H_*$. This implies that
the relevant thermodynamic potential is the Gibbs free energy of the
dual gauge theory, as opposed to the Helmholtz free energy. In our
holographic setup, the Gibbs free energy is given by the Legendre
transform of the ``on-shell" action with respect to the magnitude of
the Ramond-Ramond two-form at $r_*$, $J(r_*)$. To calculate the free
energy, we have developed an appropriate holographic renormalisation
scheme. This scheme is designed to facilitate the comparison of our
results to the results of ref.~\cite{Filev:2007gb, Erdmenger:2007bn,
  Albash:2007bk} obtained in the quenched approximation. To this end
we regularize the contribution to the free energy which due to the DBI
term of the smeared flavour branes by the addition of appropriate
covariant counterterms. The remaining contributions to the
``on-shell'' action are regularized by subtraction, using the supersymmetric solution corresponding to the limit of a vanishing external magnetic field \cite{arXiv:0807.0298} as a reference background. 
\paragraph{} 
The application of
our regularization scheme to the supersymmetric background \cite{arXiv:0807.0298} implies the identification of the bare mass parameter of the fundamental fields $m$ with the leading order coefficient in the large $r_*$ expansion of the profile of the fiducial embedding $\chi(r)$, namely $\chi(r_*)=2m/r_*+\dots$. In the limit $\epsilon_*\to 0$ ($r_*\to\infty$), this definition of the bare mass parameter $m$ agrees with the definition in the quenched approximation given in ref.~\cite{Karch:2002sh} . Moreover we define the fundamental condensate of the theory as the derivative of the Gibbs free energy with respect to the bare mass. At first order in $\epsilon_*$, our results agree with the results obtained in the quenched approximation.  At next order in  $N_f/N_c$, the relevant perturbative parameter is the IR parameter $\epsilon_q$ defined via $\epsilon_q\equiv e^{\Phi(r_q)}/e^{\Phi_*}\epsilon_*\,$. Note that $\epsilon_q$ takes into account the running of the effective 't Hooft coupling, encoded in the running of the dilaton field\footnote{Note also that to first order in $\epsilon_*$, we have $\epsilon_q=\epsilon_*+O\left(\epsilon_*^2\right).$ }. Our results for the second order in $\epsilon_q$ contribution to the fundamental condensate show that the effect of magnetic catalysis is enhanced. Furthermore we observe a logarithmic running of the condensate as a function of the UV cutoff of the theory, reflecting the positive $\beta$-function of the theory. This is the main result of our study of the effect of magnetic catalysis beyond the quenched approximation.
\paragraph{} 
An obvious direction for future studies is to investigate the effect
of finite temperature. From supergravity point of view, this
corresponds to generalizing our perturbative construction by
substituting the AdS$_5\times S^5$ zeroth order supersymmetric
background with the AdS$_5\times S^5$ black hole. Such a study would
enable us to study the phase diagram of the theory, 
which in the quenched approximation was analyzed in refs.~\cite{Erdmenger:2007bn, Albash:2007bk}, beyond the quenched
approximation. It would also be interesting to generalize our
holographic setup by turning on a finite chemical potential and to  investigate the effect of internal fundamental loops to the quantum critical points reported in refs.~\cite{Evans:2010iy, Jensen:2010vd}. 
\paragraph{} 
Finally it would be interesting to generalize our holographic setup to
other Dp/Dq--brane intersections and to apply our framework to related
phenomena also taking place at finite magnetic field. For example, it
would be interesting to construct the analogue of the effect of
Inverse Magnetic Catalysis \cite{Preis:2010cq}, observed in the
Sakai-Sugimoto Model, for holographic gauge theories dual to the
Dp/Dq--intersections and to take into account the backreaction of the
flavour branes. Other effects of potential interest include the Chiral
Magnetic Effect \cite{Kharzeev:2007jp}, 
investigated holographically in  refs.~\cite{Rebhan:2009vc, Gorsky:2010xu},  
 and the recently proposed effect of  superconductivity from $\rho$
 meson condensation in
 the QCD vacuum in a strong magnetic field \cite{Chernodub:2010qx}, 
investigated holographically in  refs.~\cite{Callebaut:2011ab,Ammon:2011je}.

\section{Acknowledgments}

We would like to thank M.~Ammon, F.~Bigazzi, A.~L.~Cotrone, K.~Y.~Kim, D.~O'Connor, A.~Paredes, R.~Rashkov, J.~Shock and J.~Tarrio for useful comments and suggestions. Moreover we would like to thank J.~Shock for critically reading the manuscript. D.~Z.~is funded by the FCT fellowship SFRH/BPD/62888/2009. Centro de F\'{i}sica do Porto is partially funded by FCT through the projects PTDC/FIS/099293/2008 \& CERN/FP/109306/2009. The work of V.~F.~is funded by an INSPIRE IRCSET-Marie Curie International Mobility Fellowship. V.~F.~would like to thank the organizers of the GGI Workshop ``Large-N Gauge Theories" in Florence and the Institute for Theoretical Physics,
at Vienna University of Technology for hospitality during the early
stages of this project. D.~Z.~thanks the gauge/gravity duality group at
the Max-Planck-Institut for Physics in Munich for hospitality
during the final stages of this work. The work of J.~E.~is partially
funded by the `Excellence Cluster for Fundamental Physics: Origin and
Structure of the Universe'.
%%%%%%%%%%%%%%%%%%%%%%%%%%%%%%%%%%%%%%%%%%%%%%%%%%%%%%%
%%%%%%%%%%%%%%%%%%%%%%%%%%%%%%%%%%%%%%%%%%%%%%%%%%%%%%%
%%%%%%%%%%%%%%%%%%%%%%%%%%%%%%%%%%%%%%%%%%%%%%%%%%%%%%%

\appendix

\section{Analytic expressions for the calculation of the EOM and the Free Energy}
\label{nonsusy-ingredients}

The analytic expressions for the sources appearing in \eqref{EOMs} and \eqref{EOM-eta} are 
\begin{eqnarray}
J_{\lambda_1}(\tilde{r}) \, & = &  \, 
\frac{4 \, {\tilde \xi}(\tilde{r}) }{{\tilde r}^4 \sqrt{{\tilde r}^4 \, + \, 1}} \, , 
\quad \quad 
J_{b_1}(\tilde{r}) \,  =  \, -\frac{8 \, {\tilde\xi}(\tilde{r})}{{\tilde r}^4 \sqrt{{\tilde r}^4 \, + \, 1}}  \, , 
\nonumber \\ \nonumber \\ 
J_{\Phi_1}(\tilde{r}) \, & = & \, 8 \, {\tilde \xi}(\tilde{r}) \, \, 
\frac{{\tilde r}^4 \,+\,\frac{1}{2}}{{\tilde r}^4 \sqrt{{\tilde r}^4 \, + \, 1}} \, , 
\quad \quad 
J_{\Lambda_1}(\tilde{r}) \,  =  \, - \, 8 \,  {\tilde \xi}(\tilde{r}) \, 
\frac{\sqrt{1+\tilde{r}^4}}{\tilde{r}^4} \, 
\frac{1 \, + \, \frac{\tilde{r}^2}{8} \, \chi_{0}'(\tilde{r})^2}{1 \, + \, \frac{\tilde{r}^2}{4} \, \chi_{0}'(\tilde{r})^2} \, ,
\nonumber \\ \nonumber \\
J_{\eta_1}(\tilde{r}) \, &  =  & \, \frac{\sqrt{{\tilde r}^4 \, + \, 1}}{{\tilde r}^3} \, \frac{ {\tilde \xi}(\tilde{r})}{ 1 \, + \, \frac{\tilde{r}^2}{4} \, \chi_{0}'(\tilde{r})^2 } \, , 
\quad \quad 
J_{H_1}(\tilde{r}) \,  =  \, \frac{8 \, \tilde\xi(\tilde{r})}{\sqrt{{\tilde r}^4 \, + \, 1}} \, ,  
\nonumber \\ \nonumber \\
J_{\Delta_1}(\tilde{r}) \, & = & \, {\tilde \xi}(\tilde{r}) \, \frac{\sqrt{1+\tilde{r}^4}}{\tilde{r}^4} \,   
\Bigg[ \, \frac{1}{{1 \, + \, \frac{\tilde{r}^2}{4} \, \chi_{0}'(\tilde{r})^2} } \, - \, 
\frac{3}{2} \, \left( 1 \, - \, \frac{5}{3}\, \cos \chi_{0}(\tilde{r})  \right) \Bigg]  \, , 
\label{nonsusyJ}
\end{eqnarray}
with 
\begin{equation}
{\tilde\xi}(\tilde{r}) \equiv \cos^3 \frac{\chi_{0}(\tilde{r})}{2} \, \sqrt{1 \, + \, \frac{\tilde{r}^2}{4} \, \chi_{0}'(\tilde{r})^2} \, . \label{xi}
\end{equation}

%%%%%%%%%%%%%%%%%%%%%%%%%%%%%%%%%%%%%%%%%%%%%%%%%%%%%%%%%%%%%%%%

\section{Useful ingredients from the massive supersymmetric solution}
\label{app-susy}

The massive supersymmetric solution was found in \cite{arXiv:0807.0298,arXiv:0909.2865} and it is easy to 
check that the following set of first order differential equations 
\begin{eqnarray}
&&
\partial_{\sigma} h \, = \, - \, Q_c \, , 
\quad \qquad
\partial_{\sigma} F \, = \, S^4 \, F 
\Bigg[ 3-2 \, \frac{F^2}{S^2} - \frac{Q_f}{2} \, e^\Phi \, \cos^4\frac{\chi_{q}}{2} \, \Bigg] \, , 
\label{massiveBPS}
\\ \nonumber \\
&& 
\partial_{\sigma} S \, = \, S^3 \, F^2 \, , 
\quad\qquad
\partial_{\sigma} \chi_{q} = -2 \, S^4 \, \tan \frac{\chi_{q}}{2} \, ,
\quad\quad
\partial_{\sigma} \Phi \, = \, Q_f \, S^4 \, e^\Phi \cos^4\frac{\chi_{q}}{2} \, .
\nonumber 
\end{eqnarray}
together with $b\, =\, 1$ \& $H\, = \, 0$ solve the full set of equations of motion, 
\eqref{EOM-chi} \& \eqref{diff-b} - \eqref{diff-H}. Changing coordinates from $\sigma$ to $\rho$, 
according to $d \rho =  S^4 \, d \sigma$, we can integrate immediately the equation of motion 
for the embedding
\begin{equation}
\sin\frac{\chi_{wv}}{2} \, = \, {e^{\rho_q} \over e^{\rho}}
\label{susyembedding-rho}
\end{equation}
where $\rho_q$ is an integration constant related to the bare quark mass. Substituting 
\eqref{susyembedding-rho} in \eqref{massiveBPS} we obtain the following set of 
explicit expressions for $S$, $F$, $\Phi$ \& $h$ for $\rho > \rho_q$
\begin{eqnarray}
&& 
S \, =  \, e^\rho \, \left[1 \, + \, \epsilon_* \, A_{S} \right]^\frac{1}{6} \, , 
\qquad \qquad \quad  
F \, = \, e^\rho\, \ ,
\frac{\left[1 \, + \, \epsilon_* \, A_{\Phi}\right]^{\frac{1}{2}}}{\left[1 \, + \, \epsilon_* \, A_{S} \right]^\frac{1}{3}} \,
\label{susysolmassive}
\\ \nonumber \\
 && 
 \Phi \, = \, \Phi_* \, - \, \log \left[ 1\, + \, \epsilon_* \, A_{\Phi} \right] \, , 
 \qquad
\partial_{\rho} h \,  = \,  - \, Q_c \, e^{-4\rho} \left[1 \, + \, \epsilon_* \, A_{S} \right]^\frac{2}{3}\ ,
\nonumber
\end{eqnarray}
with 
\begin{eqnarray}
A_{\Phi} \, & \equiv &  \, \rho_*-\rho-e^{2\rho_q-2\rho} \, + \, \frac{1}{4} \, e^{4\rho_q-4\rho} \, 
+ \, e^{2\rho_q-2\rho_*} \, - \,  \frac{1}{4} \,  e^{4\rho_q-4\rho_*}\ ,
\\ \nonumber \\ 
A_{S} \, & \equiv &  \,  \frac{1}{6} \, + \, \rho_* \, - \, \rho \, - \, \frac{1}{6} \, e^{6\rho_q-6\rho} \, 
- \, \frac{3}{2} \,  e^{2\rho_q-2\rho} \, + \, \frac{3}{4} \, e^{4\rho_q-4\rho} \, 
- \, \frac{1}{4} \, e^{4\rho_q-4\rho_*} \, + \, e^{2\rho_q-2\rho_*} \, \ ,
\nonumber
\end{eqnarray}
where $\rho_*$ is the UV scale where the dilaton takes the value $\Phi_*$ and 
$\epsilon_*=Q_f\,e^{\Phi_*}$.
Following the analysis of the main part of the paper, we redefine the radial variable in such a 
way that the warp factor keeps the standard $AdS$ form
\begin{equation}
h \, = \, \frac{R^4}{r^4} \,
\qquad \text{with} \qquad  
R^4\equiv \frac{1}{4} Q_c\ ,
\end{equation}
and expand $\partial_{\rho} h$ from \eqref{susysolmassive} in powers of $\epsilon_*$. 
Integrating, we obtain an expression for $r$ as a function of $\rho$, 
while we fix the integration constant by requiring $r(\rho_*) \, \equiv \, r_*\,  = \, e^{\rho_*}$.
Inverting that relation we have
\begin{equation}
\rho \, = \, \log r \, + \, \epsilon_* \, \rho_1 \, , 
\label{rho0}
\end{equation}
with
\begin{eqnarray}
&&
\rho_1 \, = \, \frac{1}{720} \Bigg[ 120 \log \frac{r}{r_*} \, 
+ \, 10 \left(1 - \frac{r^4}{r_*^4}\right) \left(1 - 3 \, \frac{m_0^4}{r^4}\right)\,
+\, 8 \, \frac{m_0^6}{r^6} \, \left(1 - \frac{r^{10}}{r_*^{10}}\right)
\label{rho1}
\\ \nonumber \\
&& \qquad \qquad \qquad \qquad  \qquad \qquad \,\,\,\,
+ \, 120 \, \, \frac{m_0^2}{r^2} \,  \left(1 - \frac{r^2}{r_*^2}\right) \, 
- \, 15 \, \, \frac{m_0^4}{r^4} \left(1 \, - \, \frac{r^8}{r_*^8} \right) \Bigg] \, . 
\nonumber
\end{eqnarray}
Substituting \eqref{rho0} and \eqref{rho1} in \eqref{susysolmassive} we have
\begin{eqnarray} 
&& 
S \, = \, r \left(1 \, + \, \epsilon_* \, S_1 \right) \, , 
\quad
F\, = \, r \left( 1 \, + \, \epsilon_* \, F_1 \right)  \, , 
\nonumber \\ \nonumber \\
&& 
\Phi \, = \, \Phi_* \, + \,  \epsilon_* \, \Phi_1 \, ,
\quad \& \quad 
\chi_q \,  = \, 2\, \arcsin \frac{m_0}{r} \, + \epsilon_* \chi_1 \, ,
\end{eqnarray}
with
\begin{eqnarray}
F_1
&=& - \, \frac{1}{24} \left(1 + \frac{1}{3} \,  \frac{r^4}{r_*^4} \right) \, 
+  \, \frac{1}{6} \, \frac{m_0^2}{r^2} \, - \, \frac{3}{16} \, \frac{m_0^4}{r^4}
\left(1 - \frac{1}{9} \,  \frac{r^8}{r_*^8} \right) \, +
\frac{1}{15} \,  \frac{m_0^6}{r^6}
\left(1 - \frac{1}{6} \,  \frac{r^{10}}{r_*^{10}} \right) \,,
\\ \nonumber \\
S_1
&=& \frac{1}{24} \left(1 - \frac{1}{3} \,  \frac{r^4}{r_*^4} \right) \, 
-  \, \frac{1}{12} \, \frac{m_0^2}{r^2} \, + \, \frac{1}{16} \, \frac{m_0^4}{r^4}
\left(1 + \frac{1}{3} \,  \frac{r^8}{r_*^8} \right) \, - 
\frac{1}{60} \,  \frac{m_0^6}{r^6}
\left(1 + \frac{2}{3} \,  \frac{r^{10}}{r_*^{10}} \right) \,,
\\ \nonumber \\
\Phi_1
&=& \log\frac{r}{r_*} \, + \, \frac{m_0^2}{r^2} \left(1\, - \, \frac{r^2}{r_*^2}\right) \, 
- \, \frac{m_0^4}{4 \, r^4} \left(1\, - \, \frac{r^4}{r_*^4}\right) \, ,
\\ \nonumber \\
\chi_1
&=&  \frac{1}{360} \, \frac{m_0}{\sqrt{r^2 -m_0^2}} \Bigg[ -120 \, \ln \frac{r}{m_0} \, 
+ \, 8 \left(1 \, - \, \frac{m_0^6}{r^6} \right) \, 
- \, 45 \left(1 \, - \, \frac{m_0^4}{r^4} \right) 
\nonumber \\ \nonumber \\
 && \qquad \quad \quad \, \, \, \, \, \, 
+ \, 120 \left(1 \, - \, \frac{m_0^2}{r^2} \right) \, 
+  \, \frac{r^4}{r_0^4} \left(1\, - \, \frac{m_0^4}{r^4}\right)
\left(10 \, - \, 15 \, \frac{m_0^4}{r_0^4} \, + \, 8 \, \frac{m_0^6}{r_0^6} \right) \Bigg] \, .
\label{susysolmassive-r}
\end{eqnarray}
Since they are needed for the analysis of the free energy we explicitly construct $\Lambda_1$ \& 
$\Delta_1$
\begin{eqnarray}
\Delta_1
&=& \, \frac{1}{12} \, \frac{m_0^6}{r^6} \, \left(1 \, - \,  \frac{r^2}{m_0^2} \right)^3 \,,
\label{susy-Delta1}
\\ \nonumber \\
\Lambda_1
&=& \frac{5}{72} \left(1 - \,  \frac{r^4}{r_0^4} \right) \, 
- \, \frac{5}{48} \, \frac{m_0^4}{r^4} \, \left(1 - \,  \frac{r^8}{r_0^8} \right) \, 
- \, \frac{1}{6} \, \frac{m_0^2}{r^2}
\left(1 - \,  \frac{m_0^2}{r^2} \right) \, 
+ \, \frac{1}{18} \,
\left(1 -  \frac{m_0^6}{r^6} \,  \frac{r^{10}}{r_0^{10}} \right) \, .
\label{susy-Lambda1}
\end{eqnarray}

\section{Weak magnetic field}\label{WMF}

It is possible to perform analytic computations at the limit of weak  magnetic field. The definition of the weak magnetic field that we will adopt is that the energy scale 
associated to the magnetic field $r_m$ is much smaller than the bare mass parameter $m$, namely $r_m\ll m_0$. In this limit to zeroth order in $\epsilon_*$ one has the following expansion for the profile of the fiducial embedding $\chi_0(r)$ and the bare mass parameter $m_0$
\begin{eqnarray}
\chi_0(r) \, &=& \, 2  \arcsin\left(\frac{r_q}{r}\right) \, - \, \frac{\sqrt{r^2-r_q^2}}{2r_q^3r^2}r_m^4 \, + \, {\cal O}\left(r_m^6\right) \,  ,
\\ 
m_0 \, &=& \, r_q-\frac{r_m^4}{4r_q^3} \, + \, {\cal O} \left(r_m^6\right )\nonumber \, .
\end{eqnarray} 
Expanding in  $r_m$ we get
\begin{equation} \label{WMF-emb}
\chi_0(r) \, =  \, \frac{2 m_0}{r}  \, + \, \frac{\frac{m_0^3}{3}\, +\, \frac{r_m^4}{2 m_0} }{r^3} \, + \, {\cal O}\left(r_m^6\right) \, 
\end{equation} 
which is the expression that we will use to calculate \eqref{Actsecweak}. Using \eqref{WMF-emb} we can evaluate the expressions 
for all the functions of the background through \eqref{semi-H1}, \eqref{semi-Phi1}, \eqref{semi-b1}, \eqref{semi-lambda1},  \eqref{semi-Delta1} \&  \eqref{semi-Lambda1}
and then perform the different integrals in \eqref{actionbulk}. After a long but straightforward calculation we obtain \eqref{Actsecweak}.

Here we give some details on obtaining the contribution from $\Lambda_1$, since all the rest can be calculated in a similar manner. 
Using dimensionless variables we have 
\begin{eqnarray}
J_{\Lambda_1}(\tilde{r}) \, & \sim & \, - \, \frac{8}{\tilde{r}^2} \, + \, \frac{12 \, \tilde{m}_0^2}{\tilde{r}^4} \,+\,  \frac{2 - 4\tilde{m}_0^4 }{\tilde{r}^6} \, 
+ \, {\cal O}(\tilde{r}^{-6}) \quad \& 
\\ \nonumber 
\int\limits_{\tilde r_q}^{\tilde r_*}d\tilde r\tilde r^5 J_{\Lambda_1}\Lambda_1 \, &  \sim &  \, - \, \frac{13 {\tilde r}_*^4 }{144} \, 
+ \, \frac{23  {\tilde m}_0^2 {\tilde r}_*^2 }{36} \, + \,  \frac{1}{4}\, \left( 1  \, - \,  6  {\tilde m}_0^4\right) \, \log \frac{{\tilde r}_*}{{\tilde m}_0} \, 
+ \, {\cal O}(\tilde{r}_*^{-1}) \, .
\end{eqnarray}

\end{document}